\documentclass[preprint,3p,times,twocolumn]{elsarticle}

\usepackage[T1]{fontenc}

\usepackage{amsthm}
\usepackage{amsmath,amssymb,graphicx}
\usepackage{physics}
\usepackage{mathrsfs}
\usepackage{algorithm}
\usepackage{algorithmic}
\usepackage{xcolor}
\usepackage{bm}
\usepackage{comment}
\usepackage{subfigure}
\usepackage[english]{babel}

\usepackage{stfloats}
\usepackage{cuted}
\usepackage{multirow}

\usepackage{amsfonts}
\usepackage{tablefootnote}
\usepackage{threeparttable}
\usepackage{booktabs}
\usepackage{array}
\renewcommand{\arraystretch}{1}
\usepackage{stfloats}

\usepackage{colortbl}
\usepackage{adjustbox}

\journal{ISA Transactions}
\usepackage{geometry}
\newgeometry{margin=1.8cm}
\begin{document}

\begin{frontmatter}

\title{Observer Switching Strategy for Enhanced State Estimation in CSTR Networks}

\author[label1]{Lisbel Bárzaga-Martell\corref{cor}}
\ead{lisbel.barzaga@utem.cl}
\cortext[cor]{corresponding author}
\affiliation[label1]{organization={
		Department of Electricity, Universidad Tecnológica Metropolitana (UTEM)},
	addressline={Av. José Pedro Alessandri 1242}, 
	city={ Santiago},
	postcode={7800002}, 
	country={Chile}
	}

\author[label2]{Francisco Ibáñez}
	\affiliation[label2]{organization={Department of Chemical and Bioprocess Engineering, Faculty of Engineering, Pontificia Universidad Católica de Chile},
		addressline={ Av. Vicuña Mackenna 4860}, 
		city={Santiago},
		postcode={7800002}, 
		country={Chile}
		}

\author[label3]{Angel L. Cedeño}
	\affiliation[label3]{organization={Department of Electrical Engineering, Universidad de Santiago de Chile},
		addressline={ Av. Victor Jara 3519}, 
		city={Santiago},
		postcode={9170124}, 
		country={Chile}
		}

\author[label1]{Maria Coronel}
\author[label2]{Francisco Concha}

\author[label4]{Norelys Aguila-Camacho}
\affiliation[label4]{organization={Department of Mechanical Engineering, University of North Florida},
		addressline={1 UNF Dr.}, 
		city={Jacksonville},
		postcode={32224}, 
		country={United States}
		}
\author[label2]{José Ricardo Pérez-Correa}

\begin{abstract}
Accurate state estimation is essential for monitoring and controlling nonlinear chemical reactors, such as continuous stirred-tank reactors (CSTRs), where limited sensor coverage and process uncertainties hinder real-time observability. This paper introduces a novel multi-observer switching framework that combines several advanced estimators—Extended Luenberger Observer (ELO), Extended Kalman Filter (EKF), Unscented Kalman Filter (UKF), Quadrature Kalman Filter (QKF), and Particle Filter (PF)—operating in parallel. At each sampling instant, a cost function based on the $L_1$ norm and Kullback–Leibler divergence selects the observer, yielding the best agreement with available measurements. The proposed architecture is validated through simulations of both linearized and fully nonlinear CSTR models with up to three reactors in series. Results show that the switching strategy significantly reduces estimation errors compared to single-observer approaches, especially under partial observability and parametric uncertainty. Despite its modular design, the framework remains computationally tractable, making it suitable for real-time industrial applications such as fault detection and model-based predictive control.

\end{abstract}


\end{frontmatter}

\section{Introduction}
\label{sec1}

State estimation plays a fundamental role in the safe and efficient operation of chemical processes, especially in nonlinear dynamic systems such as continuous stirred-tank reactors (CSTRs). In these systems, key variables like concentrations and reaction rates often cannot be measured directly or continuously due to sensor limitations, high costs, or harsh operating conditions \cite{kamen2012introduction, anderson2005optimal}. As a result, robust and reliable state estimation methods are essential to infer unmeasured states in real-time, enabling advanced monitoring, control, and fault diagnosis strategies.

CSTRs are commonly used in various industrial applications, including chemical synthesis, fermentation, polymerization, and wastewater treatment. Nonlinear mass and energy balances govern their dynamic behavior and are often affected by process uncertainties such as fluctuating feed conditions, model parameter variability, and unmodeled disturbances \cite{cortes_design_2021}. Limited sensor placement further amplifies the estimation challenge in practical multi-reactor configurations. While temperatures are typically measured throughout the reactor train, concentration measurements are often available only at the final stage, if at all. This partial observability introduces significant difficulties for traditional observer design, as it restricts the information available for state reconstruction.

Various nonlinear observers have been proposed and implemented in the literature to address the challenges above. The Extended Kalman Filter (EKF) \cite{grewal2014kalman} is one of the most widely used tools, offering recursive state estimation through linearization of the system model around the current state estimate. The Unscented Kalman Filter (UKF) \cite{julier1997new,farsi_state_2019,cortes_design_2021} and Quadrature Kalman Filter (QKF) \cite{chen_applying_2017,arasaratnam2007discrete} improve upon EKF by using deterministic sampling or numerical integration to better capture nonlinearities without requiring Jacobian matrices. The Particle Filter (PF) \cite{gordon1993novel}, a fully probabilistic method based on sequential Monte Carlo techniques, is suitable for strongly nonlinear and non-Gaussian systems, though it can be computationally intensive. Deterministic approaches such as Extended Luenberger Observers (ELO) \cite{Zeitz1987,sanchez_torres_robust_2016}, High-gain observers \cite{medjebouri_extended_2023}, and Nonlinear reduced-order state observers \cite{ling_state_2016} have also been applied to CSTRs, offering simpler implementations and fast convergence under ideal conditions.

Table \ref{tab:observers} synthesizes the state-observer literature on CSTRs published in the past decade. Its first column lists the observer architecture(s), the second indicates whether the design relies on a linear or nonlinear model, and the third identifies the estimated states.
\begin{table*}[ht]
\centering
\small
\caption{Overview of observers for CSTR over the past 10 years.}
\label{tab:observers}
\renewcommand{\arraystretch}{1.25}
\begin{adjustbox}{max width=\textwidth}
\begin{tabular}{|m{4.8cm}|>{\centering\arraybackslash}m{2.5cm}|>{\centering\arraybackslash}m{1.5cm}|m{4.0cm}|>{\centering\arraybackslash}m{1.0cm}|}
\hline
\textbf{Type of Observer} & \textbf{Model (CSTR)} & \textbf{\# Reactors} & \textbf{Estimated Variables} & \textbf{Ref.} \\ \hline
Fractional neural networks (FNNs) & Nonlinear  & 1 & $C_{A}, T_{1}$ & \cite{kumar_system_2024} \\ \hline
Extended Luenberger observer (ELO) & Nonlinear  & 1 & Unknown disturbance & \cite{sanchez_torres_robust_2016} \\ \hline
Extended states observer (ESO) & Nonlinear   & 1 & $C_{A}, T_{1}$ & \cite{medjebouri_extended_2023} \\ \hline
\begin{tabular}[c]{@{}l@{}}Luenberger Observer (LO),\\ Unknown Input Observer (UIO),\\ Kalman Filter (KF),\\ Sliding Mode Observer (SMO)\end{tabular} & Nonlinear & 1 & $C_{A}, T_{1}$ & \cite{cortes_design_2021} \\ \hline
Neural networks Observer (NNO) & Nonlinear   & 2 & $C_{A}, C_{B}$ & \cite{li_neural-network_2021} \\ \hline
\begin{tabular}[c]{@{}l@{}}Kalman Filter (KF),\\ Extended Kalman Filter (EKF)\end{tabular} & Nonlinear & 1 & $C_{A}, C_{B}$ & \cite{farsi_state_2019} \\ \hline
Adaptive Observer (AO) & Linear  & 1 & $C_{A},$ Sensor faults & \cite{Bzioui_adaptive_2021} \\ \hline
Cubature Kalman Filter (QKF) & Nonlinear & 1 & $C_{A}, T_{1}$ & \cite{chen_applying_2017} \\ \hline
Asymptotic Observer (AO) & Nonlinear & 1 & $C_{B}, C_{C}, k_{0,1}, k_{0,2}$ & \cite{zhao_reaction_2016} \\ \hline
Extended Kalman Filter (EKF) & Nonlinear & 1 & Model uncertainties & \cite{subramanian_adaptive_2015} \\ \hline
High-gain reduced-order observer (HGO) & Nonlinear & 1 & Model uncertainties & \cite{romero-bustamante_robust_2017} \\ \hline
\begin{tabular}[c]{@{}l@{}}Luenberger fuzzy observer (LFO),\\ Walcott-Zak fuzzy observer,\\ Luenberger fuzzy with SMC,\\ Utkin fuzzy observer\end{tabular} & Nonlinear & 1 & $C_{A}, C_{B}$ & \cite{ballesteros-moncada_fuzzy_2015} \\ \hline
\begin{tabular}[c]{@{}l@{}}Nonlinear Unknown Input\\ Observer (NUIO)\end{tabular} & Nonlinear  & 1 & $C_{A}, C_{B}, T_{1}$ & \cite{zarei_robust_2014} \\ \hline
\end{tabular}
\end{adjustbox}
\end{table*}

However, a recurring limitation in most prior work is the reliance on a single observer structure. While each observer type has known strengths, it also carries tradeoffs related to model mismatch sensitivity, noise robustness, convergence rate, and computational burden. In real-world applications where operating conditions and measurement noise levels vary dynamically, no single observer can be expected to perform optimally across all scenarios. Moreover, most reported implementations focus on single-reactor systems, and the extension to multi-stage configurations with partial state measurements remains underexplored.

Hybrid and switching strategies have emerged as promising alternatives to improve estimation reliability under such constraints. These methods leverage multiple observers or models simultaneously by blending their outputs (soft switching) or selecting the most reliable one at each time step (hard switching). Prior studies have demonstrated performance improvements in power systems, induction motors, and robotics. For instance, Ahrens and Khalil \cite{ahrens_high-gain_2009} proposed a switching gain strategy for high-gain observers to balance robustness and responsiveness. At the same time, Barut et al. \cite{barut_switching_2007} implemented a time-based switching between EKFs for parameter estimation in motor drives. In chemical processes, however, switching observers remains relatively rare. A notable exception is the work by Farsi et al. \cite{farsi_state_2019}, who applied a switched Kalman filter approach to a two-reactor CSTR system, dynamically alternating between a Linear Kalman Filter and an EKF based on the local degree of nonlinearity. While effectively reducing computational overhead, their method was restricted to a single observer type and a two-stage reactor configuration. 

On the other hand, extensive research in control engineering has leveraged switching to improve closed-loop performance, as demonstrated by Switched Fractional-Order PI/PID regulators \cite{Aguila_etal_2025, Afghoul2017} and Switched Fractional-Order Model Reference Adaptive Control schemes \cite{Aguila_Gallegos_2024}. Switching observers have emerged as a practical approach for monitoring nonlinear chemical processes. These observers typically consist of a bank of local observers and a supervisor who switches between them based on operating conditions \cite{koumboulis_switching_2012,koumboulis_switching_2017}. The design process involves defining appropriate operating areas, often using experimental data or identification techniques \cite{koumboulis_switching_2012}. Switching observers can be constructed using Lyapunov stability theory for uncertain nonlinear systems, ensuring asymptotic stability with a finite number of switches \cite{LiuSwitching}. In process monitoring applications, nonlinear observers with state-dependent gains can be designed by solving systems of partial differential equations, allowing for accurate estimation of key process variables in continuous and batch operations \cite{kazantzis_nonlinear_2000}. These techniques have been successfully applied to various chemical processes, including double-effect evaporators, catalytic batch reactors, and alcoholic fermentation processes \cite{koumboulis_switching_2012,koumboulis_switching_2017,kazantzis_nonlinear_2000}.

In this context, the present work introduces a \textbf{novel multi-observer switching framework} for enhanced state estimation in nonlinear CSTR systems. The key idea is to run a bank of heterogeneous observers in parallel, specifically, EKF, UKF, QKF, PF, and ELO, and to dynamically select the observer with the most accurate output at each sampling instant. The selection is governed by a cost function that combines two complementary error metrics: the $L_1$ norm of the estimation error and the Kullback–Leibler divergence between the measured and estimated output distributions. This composite criterion allows the switching framework to account for both the magnitude and probabilistic structure of the estimation error, thereby improving robustness to noise and unmodeled dynamics.

The proposed framework is tested on a series of increasingly complex CSTR configurations, ranging from a single reactor to a three-reactor cascade, with nonlinear dynamics and partial observability. Both linearized and fully nonlinear models are considered, and the observer performance is evaluated using simulation experiments that include Gaussian sensor noise and parametric uncertainty. The estimation accuracy is quantified using standard performance metrics (e.g., $L_2$ norm, $L_{\infty}$ norm, MSE), and the computational cost of each method is also reported. In addition, Monte Carlo simulations are used to assess the robustness of the switching strategy under random variations in model parameters such as reaction rates, activation energy, and heat transfer coefficients. Series reactors are widely used in industries like pharmaceuticals and biotechnology due to their suitability for advanced control. Prior studies have shown their real-time monitoring and disturbance rejection effectiveness using observer-based strategies \cite{sainz-garcia_adaptive_2022, ling_state_2016}.

The main contributions of this study are:
\begin{enumerate}
    \item Development of a switching observer architecture that integrates multiple nonlinear estimation methods within a unified framework for CSTR systems.
    \item Design of a cost-based switching strategy, incorporating both deterministic ($L_1$ norm) and probabilistic (Kullback–Leibler divergence) error metrics for observer selection.
    \item Application to multi-stage nonlinear reactor networks with limited sensor information, demonstrating scalability and practical relevance to industrial scenarios.
    \item Comprehensive simulation-based evaluation, including performance benchmarking, noise resilience, and parametric robustness through Monte Carlo analysis.
\end{enumerate}

This work advances the state of the art in nonlinear observer design for chemical process systems and offers a viable pathway for real-time deployment in applications such as soft sensing, fault detection, and model predictive control.
\begin{figure*}[t]
 \centering
\includegraphics[width=0.8\linewidth]{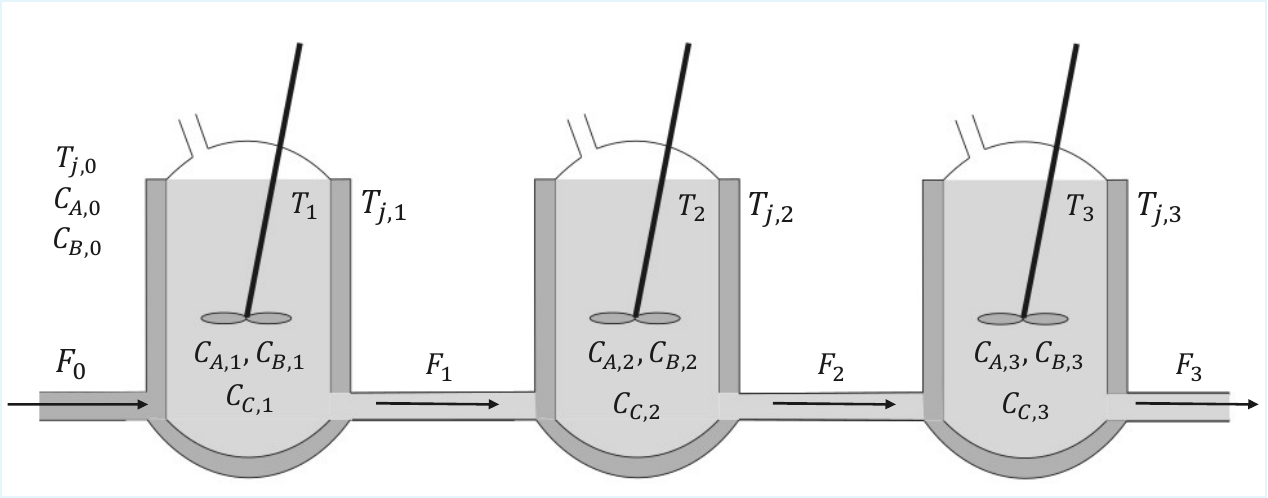} 
  \caption{Diagram of the CSTR process.}
 \label{fig1}
\end{figure*} 

The remainder of the paper is organized as follows. Section~\ref{sec2} introduces the nonlinear and linearized models of the three-tank system. Section~\ref{sec3} details the observers and the switching algorithm. Simulation results are presented in Section~\ref{sec4}, and conclusions are drawn in Section~\ref{sec5}.

\section{Continuous Stirred-Tank Reactor (CSTR) with interchange: description and modeling}
\label{sec2}

This section presents the general description of a 3-tank CSTR process and its corresponding model, which will be used for observer design and simulation in the subsequent sections. Based on mass and energy balances, nonlinear and linearized models of the CSTR were developed.

\subsection{Process description}

A CSTR is a type of reactor in which substances react continuously under constant stirring, ensuring uniform conditions within the reactor. The CSTR operates with a continuous inflow of reactants and outflow of products, maintaining a well-mixed system. This setup implies that concentrations and temperatures remain homogeneous in the reactor, providing consistent reaction conditions and simplifying the control of reaction variables. An exothermic reaction sequence \(A + B\rightarrow C\) occurs in the reactor.

\subsection{Nonlinear model of the CSTR process}

Figure \ref{fig1} illustrates the process to be addressed in this work, corresponding to three continuous stirred tank reactors (CSTR) arranged in series, designed to perform an irreversible elementary reaction in the liquid phase. Reactants \(A\) and \(B\) enter the first reactor with a constant volumetric flow \(F_{0} \ [\mathrm{L}/\mathrm{min}]\) and initial concentrations \(C_{A,0} \ [\mathrm{mol}/\mathrm{L}]\) and \(C_{B,0} \ [\mathrm{mol}/\mathrm{L}]\). Each reactor has an operating volume of \(V = 100 \ [\mathrm{L]}\), and at the beginning of the reaction, all reactors are filled with an inert solvent. Furthermore, density variations due to the chemical reaction are assumed to be insignificant and do not affect the overall balance. The temperatures in the reactors ($T_{i=1 \ldots 3}, \  [\mathrm{K}]$) are regulated by cooling jackets, which ensure a constant fluid flow. The cooling jacket temperature ($T_{j,0} \ [\mathrm{K}]$) is the same for all three reactors. Finally, the physical properties of the reactants, such as density and specific heat (\(c_p\)), are assumed to remain constant throughout the process.

The nonlinear CSTR model includes five state variables for each reactor: the concentration of reactant A, (\(C_{A}\)), concentration of reactant B, (\(C_{B}\)), concentration of reactant C, (\(C_{C}\)), reactor internal temperature, (\(T\)) and jacket temperature, (\(T_{j}\)). The energy and mass balance equations are described below.

\textbf{Mass balances}

\textit{ First reactor}
\begin{align}
\frac{dC_{A,1}}{dt} &= \frac{F_0}{V} C_{A,0} - \frac{F_1}{V} C_{A,1} - k_1 C_{A,1} C_{B,1},\label{eq1a}\\
\frac{dC_{B,1}}{dt} &= \frac{F_0}{V} C_{B,0} - \frac{F_1}{V} C_{B,1} - k_1 C_{A,1} C_{B,1},\label{eq1b}\\
\frac{dC_{C,1}}{dt} &= k_1 C_{A,1} C_{B,1} - \frac{F_1}{V} C_{C,1}\label{eq1c}.
\end{align}

\textit{Reactors ($i=2,3$)}
\begin{align}
\frac{dC_{A,i}}{dt} &= \frac{F_i}{V} (C_{A,i-1} - C_{A,i}) - k_i C_{A,i} C_{B,i},\label{eq2a}\\
\frac{dC_{B,i}}{dt} &= \frac{F_i}{V} (C_{B,i-1} - C_{B,i}) - k_i C_{A,i} C_{B,i}, \label{eq2b} \\  
\frac{dC_{C,i}}{dt} &= k_i C_{A,i} C_{B,i} - \frac{F_i}{V} C_{C,i},\label{eq2c}
\end{align}
where $C_{A,i}, C_{B,i}$, and $C_{C,i}$ represent the concentrations of compound A, compound B, and compound C, respectively. $C_{A,0}$ and $C_{B,0}$ are the concentrations at the inlet of reactor 1. $F_{0}$ is the inlet flow, $F_{1}$, $F_{2}$ and $F_3$ are the output flows of reactors 1, 2, and 3, respectively, and $V$ is the volume of reactors.

The kinetic parameters of the chemical reactions occurring within the reactor are defined as:
\begin{equation}
k_i = k_0 \exp \left( \frac{-E}{R_{g}T_i} \right); i=1 \ldots 3,
\end{equation}
where $k_{0}$ is the pre-exponential factor, $E$ is the activation energy, and $R$ is the universal gas constant.

\textbf{Energy balances}: Each reactor temperature \(T_i\), $i=1\ldots 3$ is modeled considering heat transfer and the heat of reaction as
\begin{align}
\label{eq3a}
\frac{dT_i}{dt} &= \frac{F_i}{V}(T_{i-1} - T_i) - \frac{\Delta H}{\rho c_p} k_i C_{A,i} C_{B,i} + \frac{UA}{\rho c_p V} (T_{j,i} - T_i),
\end{align}
while the temperature of the jacket \(T_{j,i}\), $i=1\ldots 3$ is governed by
\begin{equation}
\label{eq4a}
\frac{dT_{j,i}}{dt} = \frac{F_j}{V_j}(T_{j,0} - T_{j,i}) - \frac{UA}{\rho_j c_{pj} V_{j}}(T_{j,i} - T_i),
\end{equation}
where $T_{i}$ and $T_{j,i}$ represent the internal and jacket temperatures in the reactors, respectively. $\Delta H$ is the reaction enthalpy, $\rho$ and $\rho_{j}$ are the mixture densities in the reactor and the jacket. $c_{p}$ and $c_{pj}$ are the reactor and jacket specific calorific capacity, respectively, and $UA$ is the heat transfer coefficient per area. $V_{j}$, $T_{j,0}$, and $F_{j}$ denote the jacket volumes, the inlet jacket temperature, and the jacket flow, respectively, and they are considered the same in each reactor.

The energy and mass balances can be represented in a nonlinear state space form (Note that the time dependence is omitted in the continuous-time equations for ease of presentation):
\begin{align}
    \label{eqfnonlinear}
\dot{\mathbf{x}} & = f(\mathbf{x,u}), \\
\label{eqhnonlinear} 
\mathbf{y} & = h(\mathbf{x}), 
\end{align}
where the state vector $\mathbf{x}$ has a different number of components depending on the number of reactors in the system. For $q$  reactors, the state vector is given by $\mathbf{x} = [\mathbf{x}^{(1)T},\dots,\mathbf{x}^{(q)T}]^{T}$, where $\mathbf{x}^{(i)} = [C_{A,i}, C_{B,i}, C_{C,i}, T_{i}, T_{j,i}]^{T}$ for $i = 1, 2, 3$ (see Table 2), the input vector is $\mathbf{u} = [F_{0} \ C_{A,0} \ C_{B,0} \ T_{0} \ T_{j,0}]^{T} \in \mathbb{R}^{5\times 1}$, and $\mathbf{y} = h(\mathbf{x}) \in \mathbb{R}^{p}$ is the output vector (see Table \ref{tab_observability}). The nonlinear function $f(\mathbf{x,u})$ is defined in \ref{funciones}. 

\subsection{Linear model equations}
One way to approach nonlinear system estimation is to transform the original nonlinear model into an approximate linear model with similar variables. To obtain the linearized model of the CSTR system, we adopt the following assumptions:

\textbf{Assumption 1 \label{A1}}: Only the first reactor model is linearized and the nonlinear function $f(\mathbf{x,u}) \in \mathbb{R}^{3\times 1}$ is continuously differentiable ($C^1$) in an open neighborhood around some $\mathbf{x}(0)$.

\textbf{Assumption 2 \label{A3}}: The internal temperature ($T_{i}$) and the jacket temperature ($T_{j,i}$) remain constant, so their effect on linearization is not appreciable. This assumption is reasonable because these temperatures are controlled variables in each reactor, so they remain at some specific set point.
 
\textbf{Assumption 3 \label{A4}}: The inlet and outlet flow are bounded: $0 \leq F_{0} \leq \bar{F}_{0}$ and $0 \leq F_{1} \leq \bar{F}_{1}$.

\textit{Assumption 1} means that we can use a first-order Taylor series approximation of the nonlinear equations obtained to linearize the nonlinear model of the first reactor of the CSTR around an operating point. The linearized state space representation will have the structure described in equations \eqref{eq5c}-\eqref{eq5d}:
\begin{align}
    \label{eq5c}
    \dot{\mathbf{x}} &= \mathbf{A}\mathbf{x}+\mathbf{B}\mathbf{u},\\
    \label{eq5d}
    y  &= \mathbf{C}\mathbf{x},
\end{align}
where $\mathbf{x} = \begin{bmatrix}  C_{A,1} & C_{B,1} & C_{C,1} \end{bmatrix}^{T} \in \mathbb{R}^{3\times 1} $ and $\mathbf{u} = \begin{bmatrix} F_{0} & C_{A,0} & C_{B,0} \\ \end{bmatrix}^{T} \in \mathbb{R}^{3\times 1}$ are the state and input vectors, respectively. The matrices $\mathbf{A}$ and $\mathbf{B}$ correspond to the Jacobians of $f(\mathbf{x},\mathbf{u})$, specifically:
\begin{equation}
    \mathbf{A} = \frac{\partial f}{\partial \mathbf{x}} \bigg|_{(\mathbf{x}(0), \mathbf{u}(0))},\qquad \mathbf{B} = \frac{\partial f}{\partial \mathbf{u}} \bigg|_{(\mathbf{x}(0), \mathbf{u}(0))},
\end{equation}
with $(\mathbf{x}(0),\mathbf{u}(0))$ the operating point around which the linearization is made. Finally, the partial derivatives lead to the linearized model of the CSTR given in \eqref{eq5c}-\eqref{eq5d} with:
\begin{align}
    \mathbf{A} &=
    \begin{bmatrix}
    -\left (\frac{F_{1}}{V} +k_{1}C_{B,1} \right ) & -k_{1}C_{A,1} & 0 \\
    -k_{1}C_{B,1} & -\left (\frac{F_{1}}{V} +k_{1}C_{A,1} \right ) & 0 \\
    k_{1}C_{B,1} & k_{1}C_{A,1} & -\frac{F_{1}}{V} \\
    \end{bmatrix},\label{eqlina}\\ 
    \mathbf{B} &=
    \begin{bmatrix}
    \frac{C_{A,0}}{V} & \frac{F_{0}}{V} & 0 \\
    \frac{C_{B,0}}{V} & 0 & \frac{F_{0}}{V} \\
    0 & 0 & 0 \\
    \end{bmatrix}, \label{eqlinb}\\
    \mathbf{C} &= 
    \begin{bmatrix} 0 & 0 & 1\end{bmatrix}\label{eqlinc}.
\end{align}

\section{Problem formulation and observer's design}
\label{sec3}

In industrial systems of cascade CSTRs, temperatures are routinely monitored in each vessel. In contrast, concentrations—critical process variables—are typically measured only in the last reactor because suitable sensors are expensive, less accessible, and often less reliable. However, for process monitoring and control purposes and improved efficiency, it is crucial to have accurate estimations of concentrations in the intermediate stages (reactors). This implies that the complexity of the estimation increases as the number of reactors in series grows. This work will address three cases separately, with increasing complexity, as follows. 

\textbf{Case 1}: The estimation considers a single reactor (Reactor 1). The measured variables are $C_{C,1}$ (outlet product concentration), $T_{1}$ (reactor temperature) and $T_{j,1}$ (outlet jacket temperature), while the estimated variables are $\hat{C}_{A,1}$ and $\hat{C}_{B,1}$. The process is modeled for simulation with five nonlinear differential equations, with a state vector given by $\mathbf{x}$ in Table \ref{tab_observability}.

\textbf{Case 2}: The estimation considers two reactors in series. The dimension of the state vector, $\mathbf{x}$, increases to 10 (see Table \ref{tab_observability}). The differential equations representing this model are given by Eq. \eqref{eq1a}-\eqref{eq1c}, Eq. \eqref{eq2a}-\eqref{eq2c} with $i=2$ and \eqref{eq3a}, \eqref{eq4a}, with $i=1,2$. All temperatures are measured, but concentrations are measured only in the last reactor (Reactor 2). The measurements vector is shown in Table \ref{tab_observability}.  $\hat{C}_{A,1}$, $\hat{C}_{B,1}$, and $\hat{C}_{C,1}$ are the estimated concentrations in Reactor 1. This scenario illustrates how to infer the behavior of upstream reactors using only downstream measurements, when early-stage sensors are technically challenging or economically infeasible. Thus, a full-state observer becomes critical to "see inside" the unmeasured part of the system.

\textbf{Case 3}: The most complex scenario considered in this study involves three reactors in series. The resulting state vector is of dimension 15 (see $\mathbf{x}$ in Table \ref{tab_observability}). The differential equations representing the system are given by Eq. \eqref{eq1a}-\eqref{eq1c}, Eq. \eqref{eq2a}-\eqref{eq2c} with $i=2,3$, and \eqref{eq3a},\eqref{eq4a}, with $i=1,2,3$. All temperatures and the concentrations of the last reactor are considered measured, see Table \ref{tab_observability}. $\hat{C}_{A,1}$, $\hat{C}_{B,1}$, $\hat{C}_{C,1}$, $\hat{C}_{A,2}$, $\hat{C}_{B,2}$, and $\hat{C}_{C,2}$ are the estimated concentrations in reactors 1 and 2. This case tries to represent the lack of instrumentation in early process stages, due to hostile environments or physical inaccessibility, such as high pressures and temperatures, or a corrosive environment. Estimating critical variables can improve process safety and efficiency.

\textbf{Special case}: Like in Case 1, a single reactor is considered here. However, the process is simulated by the linear model given in Eq. \eqref{eq5c} - \eqref{eqlinc}. Hence, linear observers can be designed. Although temperatures can be measured, they do not need to be used in the observer design.

\begin{table*}[b]
\renewcommand{\arraystretch}{1.3}
\caption{Observability analysis for the different cases addressed in this work.}
\label{tab_observability}
\centering
\begin{tabular}{|c|c|c|c|c|}
\hline
\textbf{Case} & \textbf{State Variable} & \textbf{Measured Variable} & \textbf{Variables of interest} & \textbf{Observability} \\ \hline
\textbf{SC}    
& \begin{tabular}[c]{@{}c@{}}$\mathbf{x} = [C_{A,1}, C_{B,1}, C_{C,1}]^{\top}$\end{tabular}   
& $\mathbf{y} = C_{C,1}$    
& $\hat{C}_{A,1}, \hat{C}_{B,1}$   
& FO \\ \hline

\textbf{1}    
& \begin{tabular}[c]{@{}c@{}}$\mathbf{x} = \mathbf{x}^{(1)}$\end{tabular}   
& $\mathbf{y} = [C_{C,1}, T_{1}, T_{j,1}]^{\top}$    
& $\hat{C}_{A,1},\hat{C}_{B,1}$   
& FO \\ \hline

\textbf{2}    
& \begin{tabular}[c]{@{}c@{}}$\mathbf{x} = [\mathbf{x}^{(1)\top},\mathbf{x}^{(2)\top}]^{\top}$\end{tabular}     

& \begin{tabular}[c]{@{}c@{}}$\mathbf{y} = [T_{1}, T_{j,1}, C_{A,2}, $\\ $C_{B,2}, C_{C,2}, T_{2}, T_{j,2}]^{\top}$\end{tabular}                                                           
& \begin{tabular}[c]{@{}c@{}}$\hat{C}_{A,1}, \hat{C}_{B,1}, \hat{C}_{C,1}$\end{tabular}                                                             
& PO \\ \hline

\textbf{3}    
& \begin{tabular}[c]{@{}c@{}}$\mathbf{x} = [\mathbf{x}^{(1)\top},\mathbf{x}^{(2)\top},\mathbf{x}^{(3)\top}]^{\top}$\end{tabular}       

& \begin{tabular}[c]{@{}c@{}}$\mathbf{y} = [T_{1}, T_{j,1}, T_{2}, T_{j,2},$ \\   $C_{A,3}, C_{B,3}, C_{C,3}, T_{3}, T_{j,3}]^{\top}$\end{tabular}                                                           
& \begin{tabular}[c]{@{}c@{}}$\hat{C}_{A,1}, \hat{C}_{B,1}, \hat{C}_{C,1},$\\ $\hat{C}_{A,2}, \hat{C}_{B,2}, \hat{C}_{C,2}$\end{tabular}                                                             
& PO \\ \hline
\end{tabular}
\begin{tablenotes}
\item  Note that $\mathbf{x}^{(i)} = [C_{A,i}, C_{B,i}, C_{C,i}, T_{i}, T_{j,i}]^{\top}$ for $i = 1, 2, 3$. ``SC" indicates \textbf{Special case}.``FO" indicates \textbf{Fully Observable}. ``PO" indicates \textbf{Partially Observable}.
\end{tablenotes}
\end{table*}

\subsection{Observability analysis}
\label{sec_obs}

Before designing an observer for a dynamic system, it is essential to assess its observability, as this property determines whether the system's internal states can be uniquely inferred from output measurements over time. In linear systems, observability provides an explicit criterion for the feasibility of constructing a state observer, such as a Luenberger observer or a Kalman filter. The concept becomes more complex for nonlinear systems and must often be evaluated using local or differential criteria, yet it remains equally critical. Designing an observer for an unobservable system may lead to inaccurate or even unstable state estimates, ultimately compromising the effectiveness of control or monitoring strategies. 

Therefore, verifying observability is a fundamental step in linear and nonlinear observer design, and this work must address it.

\textbf{Special case}: For the CSTR linearized system in Eq. \eqref{eq5c}-\eqref{eqlinc}, the test is simple because a necessary and sufficient condition for observability is that the observability matrix has full rank \cite{czyzniewski_observability_2022}. In the case of CSTR, the observability linear matrix $\mathcal{O}_{L}$ is composed of the matrices $\mathbf{A}$ (Eq. (\ref{eq5c})) and  $\mathbf{C}$ (Eq. (\ref{eq5d})) as:
\begin{equation}
\label{eq7a}
\mathcal{O}_{L}=\left [C \quad CA \quad CA^{2}  \right ]^{T},
\end{equation}
which has full rank (rank $=3$). The system is thus fully observable from $C_{C,1}$, which means that if only $C_{C,1}$ is measured, it is possible to completely reconstruct the states of the system (\( C_{A,1} \) and $C_{B,1}$).

\textbf{Cases 1 to 3}: For these cases, where the system is nonlinear and represented as in Eq. \eqref{eqfnonlinear}, \eqref{eqhnonlinear}, the observability can be analyzed using canonical forms, for which the invertibility of the observability map and global observability are guaranteed. The observability nonlinear matrix $\mathcal{O}_{NL}$ is defined by \cite{czyzniewski_observability_2022}:
\begin{equation}
\label{eq7b}
\mathcal{O}_{NL}=\left [\frac{\partial h(\mathbf{x})}{\partial \mathbf{x}} \quad \frac{\partial \mathcal{L}_{f}h(\mathbf{x})}{\partial \mathbf{x}}  \quad \cdots \quad \frac{\partial \mathcal{L}_{f}^{n-1}h(\mathbf{x})}{\partial \mathbf{x}}  \right ]^{T} , 
\end{equation}
where $\mathcal{L}_{f}h(\mathbf{x})$ is the first Lie derivative of $h(\mathbf{x})$ along $f(\mathbf{x},\mathbf{u})$ and the rest are the higher-order Lie derivatives.

Since functions $f(\mathbf{x},\mathbf{u})$, $h(\mathbf{x})$ are different in each case (see Table \ref{tab_observability}), the observability analysis must be performed independently. 

The analysis was performed by comparing the rank of the matrix $\mathcal{O}_{NL}$ with the dimension of the state vector for each case. In Case 1, $\text{rank}(\mathcal{O}_{NL})=5$, which is equal to the dimension of the state, indicates that the system is globally observable. In Case 2, the dimension of the state vector is 10, while $\text{rank}(\mathcal{O}_{NL})=9$. As a result, the system is partially observable. In Case 3, the dimension of the state vector is 15, and $\text{rank}(\mathcal{O}_{NL})=13$, which also indicates that the system is partially observable.

In summary, for Cases 2 and 3, it is impossible to verify global observability because the observability matrix does not have full rank. When additional reactors are connected in series, the observability analysis becomes progressively more complex because the set of state variables to be estimated and the corresponding measurements expand. Table \ref{tab_observability} presents the result of the observability analysis for each case, also detailing the defined state variables, the number of measurements, and the state variables that need to be estimated.
\subsection{Switching framework}
\label{sec31}
\begin{figure*}[ht]
 \centering
 \includegraphics[width=0.7\linewidth]{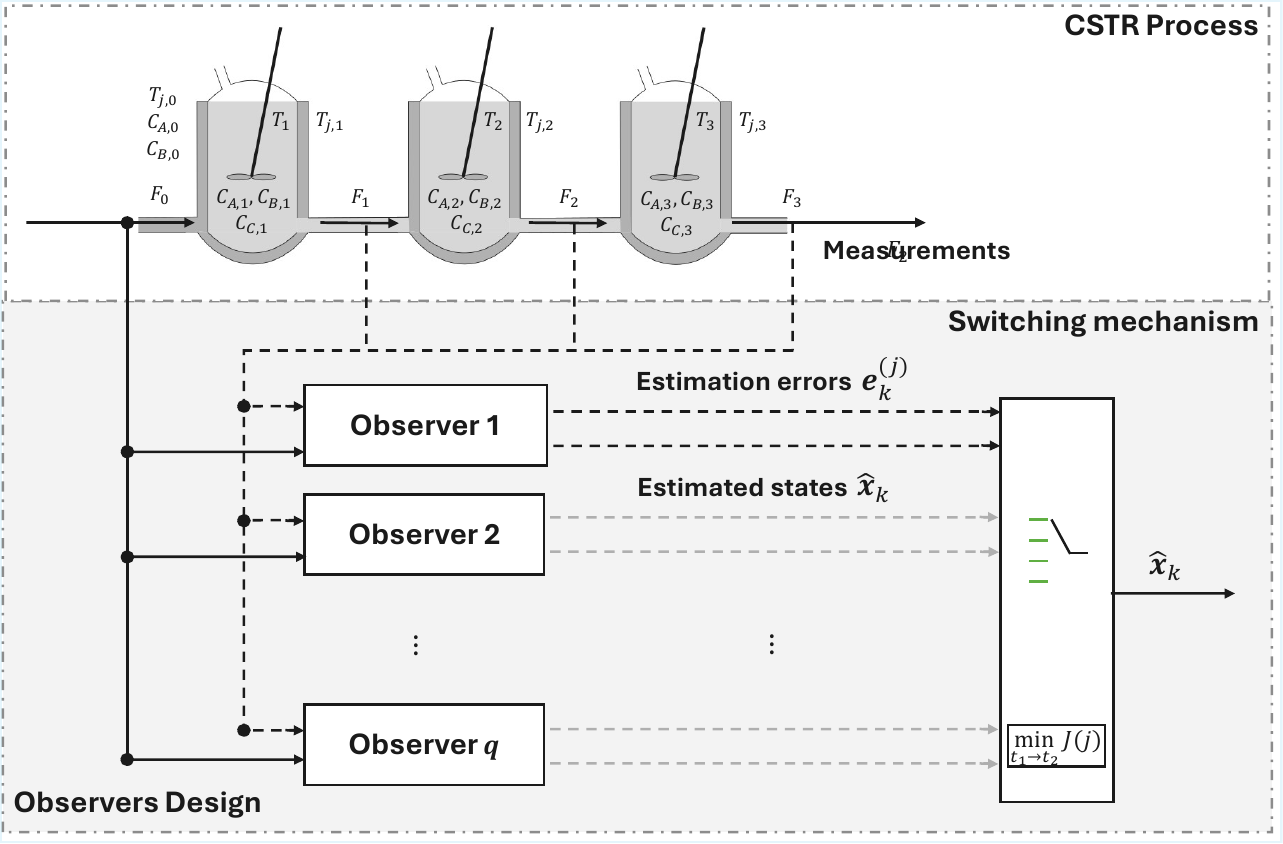} 
  \caption{Block diagram of the state observer switched approach proposed in this work for the nonlinear CSTR process.}
 \label{fig2}
\end{figure*}

\subsection{Observers Design}
\label{sec32}

The design of observers in chemical processes is challenging because of the system's high nonlinearity. This study aims to design accurate and reliable observers that explicitly consider the nonlinearities of cascade CSTR systems. Rather than using a single observer for the estimation process, we propose using \(q\) observers with various mathematical formulations, integrated into a switching framework that selects the estimated output according to a specific rule. A general diagram of the proposed approach is shown in Figure \ref{fig2}.

The proposed observers to be used in the switching approach are the Extended Luenberger observer (ELO), the Extended Kalman filter (EKF), the Quadrature Kalman filter (QKF), the Particle Filter (PF), and the Unscented Kalman filter (UKF), whose equations are described in \ref{ELO} to \ref{PF}. These observers have nonlinear and linear versions, allowing them to be used in Cases 1 to 3 and the Special Case detailed at the beginning of Section \ref{sec3}. 

All observers were designed based on discrete versions of the CSTR reactor's linear and nonlinear models, where the discretization was performed using the Euler method. The discrete models are represented in the discrete state-space nonlinear form:
\begin{align}
\label{eq9}
        \mathbf{x}_{k+1} & = f(\mathbf{x}_{k},\mathbf{u}_{k}) + \bm{\omega}_{k}, \\  \label{eq9b}
    \mathbf{y}_{k} & = h(\mathbf{x}_{k}) + \bm{\upsilon}_{k}, 
\end{align}
and for the linear case:
\begin{align}
\label{eq10}
     \mathbf{x}_{k+1} & = \mathbf{A}\mathbf{x}_{k} + \mathbf{B}\mathbf{u}_{k} + \bm{\omega}_{k}, \\ \label{eq10b}
     y_{k} & = \mathbf{C}\mathbf{x}_{k} + \bm{\upsilon}_{k}, 
\end{align}
with $k$ representing the discretized time instant, $\bm{\omega}_{k}$ and $\bm{\upsilon}_{k}$ denoting the disturbances for the models and the outputs, respectively.  

For all cases, all observers share the same input vector $\mathbf{u}$, which includes the inlet flow $F_{0}$ to Reactor 1 and the inlet concentrations $C_{A,0}$ and $C_{B,0}$. Observers also use the available measured states from each reactor as inputs, see Table \ref{tab_observability} for details. The outputs generated by the observers include the estimated state vector $\hat{\mathbf{x}}_{k}$ and the estimation error:
\begin{align}\label{eq_error}
    \mathbf{e}_{k} = \mathbf{y}_k-\hat{\mathbf{y}}_k,
\end{align}
where $\mathbf{y}_k$ is given in Table \ref{tab_observability} and $\hat{\mathbf{y}}_k$ is defined according to the case as follows:
\begin{itemize}
    \item  \textbf{Special Case:}
    \begin{equation}
    \label{eq_error0}
    \hat{y}_k=\hat{C}_{C,1},
    \end{equation}
  
    \item \textbf{Case 1 (1 reactor)}: 
    \begin{equation}
    \label{eq_error1}
    \hat{\mathbf{y}}_k=[\hat{C}_{C,1},\hat{T}_{1},\hat{T}_{j,1}]^{\top},
    \end{equation}
    \item \textbf{Case 2 (2 reactors)}:
    \begin{equation}
    \label{eq_error2}
    \hat{\mathbf{y}}_k=[\hat{T}_{1}, \hat{T}_{j,1}, \hat{C}_{A,2}, \hat{C}_{B,2}, \hat{C}_{C,2}, \hat{T}_{2}, \hat{T}_{j,2}]^{\top},
    \end{equation}
    \item \textbf{Case 3 (3 reactors)}:
    \begin{equation}
    \label{eq_error3}
    \hat{\mathbf{y}}_k=[\hat{T}_{1}, \hat{T}_{j,1}, \hat{T}_{2}, \hat{T}_{j,2},\hat{C}_{A,3}, \hat{C}_{B,3}, \hat{C}_{C,3}, \hat{T}_{3}, \hat{T}_{j,3}]^{\top},
    \end{equation}
\end{itemize}

After designing all observers, the switching rule was established. The switching framework collects the estimated states $\hat{\mathbf{x}}_{k}$ from each observer and selects one to output based on a predefined rule, as illustrated in Figure \ref{fig2}.

The switching framework developed in this work is designed to select the most reliable estimate at each sampling instant, utilizing a decision rule grounded in a normalized cost function. The cost function formulation draws upon well-established strategies reported in the literature. Specifically, numerous studies have proposed performance-driven switching laws, wherein the cost function is typically constructed based on estimation error norms, Lyapunov-based criteria, or weighted combinations of error dynamics, tailored to the characteristics of the system and the estimation objectives. A representative example of such an approach is provided by Narendra et al. \cite{narendra_adaptive_2000}, who introduced a switching strategy between competing models based on a performance index that captures recent prediction accuracy.

The cost function in this work is defined for each observer $j$ as:
\begin{equation}
J_{(j)}= \frac{\left\| \mathbf{e}_{k}^{(j)}\right\|_{1}}{\textnormal{max}\left ( \left\| \mathbf{e}_{k}^{(l)}\right\|_{1} \right )} + \frac{\left\| \mathbf{e}_{k}^{(j)}\right\|_{KL}}{\textnormal{max}\left ( \left\| \mathbf{e}_{k}^{(l)}\right\|_{KL} \right )},
\label{eq40}
\end{equation}
where $\mathbf{e}_{k}^{(j)}$ denotes the output estimation error of observer $j$, measured in both the $L_1$ norm and the Kullback-Leibler (KL) divergence. Depending on the case, vector $\mathbf{e}_{k}^{(j)}$ has different dimensions, as detailed in Section \ref{sec32}. Subindex $l$ in $\mathbf{e}_{k}^{(l)}$ \eqref{eq40} runs through the five observers. This formulation ensures that the cost assigned to each observer is relative to the worst-performing observer in each metric, allowing a balanced comparison. The observer yielding the minimum cost $J(j)$ at each sample time is selected to provide the system's estimated state.

The \( L_1 \) norm, \( \|\mathbf{e}_{k}^{(j)}\|_1 \) used in \eqref{eq40}, represents the sum of the absolute values of the error components, offering a measure of the total magnitude of the estimation error. It is defined as:
\begin{equation}
\|\mathbf{e}_{k}^{(j)}\|_1 = \sum_{i=1}^{n} \left| \mathbf{e}_{k,i}^{(j)} \right|,
\end{equation}
where \( \mathbf{e}_{k,i}^{(j)} \) denotes the \( i \)-th component of the output estimation error vector $\mathbf{e}_{k}^{(j)}\in\mathbb{R}^{n \times 1}$. The Kullback-Leibler (KL) divergence, \( \|\mathbf{e}_{k}^{(j)}\|_{\text{KL}} \), measures the information loss when the estimated state distribution is used to approximate the true distribution of \( \mathbf{x}_k \). It provides a probabilistic perspective by capturing the mismatch between the two distributions. The KL divergence is given by the following:
\begin{equation}
\|\mathbf{e}_{k}^{(j)}\|_{\text{KL}} = \sum_{i=1}^{n} \vert x_{k,i}\vert \log\left(\left |\frac{x_{k,i}}{\hat{x}_{k,i}}\right | \right),
\end{equation}
where \( x_{k,i} \) and \( \hat{x}_{k,i} \) are the \( i \)-th components of the true and estimated state vectors, respectively.

The composite cost function \eqref{eq40} integrates both error metrics to evaluate observer performance from complementary perspectives: the \( L_1 \) norm captures the overall error magnitude, while the KL divergence reflects the distributional fidelity of the estimate. Including the KL divergence ensures that the selected observer not only minimizes raw estimation error but also aligns with the statistical structure of the system’s state, thereby increasing confidence in the switching framework’s decision.

\subsection{Simulation set-up}

All simulations were conducted in MATLAB @2024 over a $40$ [min] horizon to demonstrate the performance of the switching framework introduced in Section \ref{sec31}. Table \ref{tab_observability} summarizes three operating scenarios for the nonlinear CSTR and one for the linearized CSTR.

The initial concentrations in the first reactor were set to:
$ C_{A,1}(0)=1\ [\mathrm{mol/L}],\ C_{B,1}(0)=1\ [\mathrm{mol/L}],\ C_{C,1}(0)=0\ [\mathrm{mol/L}].$ In reactors 2 and 3, the concentrations of both reactants and the product were zero: $
C_{A,i}(0)=C_{B,i}(0)=C_{C,i}(0)=0\ [\mathrm{mol/L}]\quad (i=2,3).$ These values define each case's initial state vector \(\mathbf{x}_0\). All other parameters used in the CSTR model (Equations \eqref{eq1a}–\eqref{eq4a}) are listed in Table \ref{tabla3}.

\begin{table}[ht]
\renewcommand{\arraystretch}{1.1}
\caption{CSTR Parameters values used for simulation.}
\label{tabla3}
\vspace{0.25cm}
\centering
\begin{tabular}{|c|c|c|}
\hline 
\textbf{Parameter}  & \textbf{Value} & \textbf{Units} \\
\hline
\(F_{0}\)      & $6$     & $\mathrm{L}/\mathrm{min}$ \\
\hline
\(F_i\)      & $12$     & $\mathrm{L}/\mathrm{min}$ \\
\hline
$F_{j}$             & $30$        & $\mathrm{L}/\mathrm{min}$ \\
\hline
$V$            & $100$   & $\mathrm{L}$ \\
\hline
$V_j$            &$50$   & $\mathrm{L}$ \\
\hline
$k_{0}$    & $5.0 \times 10^5$ & $\mathrm{L}/(\mathrm{mol*min})$ \\
\hline
$E$        & $1.0\times 10^4$  & $\mathrm{cal}/\mathrm{mol}$ \\
\hline
$R_{g}$        & $1.987$  & $\mathrm{cal}/\mathrm{mol}\,\mathrm{K}$ \\
\hline
$\Delta H$ & $-4.0 \times 10^4$ & $\mathrm{cal}/\mathrm{mol}$\\
\hline
$\rho, \rho_j$     &  $1.0 \times 10^3$     & $\mathrm{g}/\mathrm{L}$ \\
\hline
$c_{p}, c_{pj}$    & $4.18$      & $\mathrm{cal}/\mathrm{g}\,\mathrm{K}$ \\
\hline
$UA$        & $1.0 \times 10^5$   & $\mathrm{cal}/\mathrm{min}\,\mathrm{K}$ \\
\hline
$T_1(0)$ & $300$ & $\mathrm{K}$ \\
\hline
$T_2(0)$ & $325$ & $\mathrm{K}$ \\
\hline
$T_3(0)$ & $350$ & $\mathrm{K}$ \\
\hline
$T_{j,i}(0)$ & $370$ & $\mathrm{K}$ \\
\hline
\end{tabular}
\end{table}   

The observer's parameters for ELO, EKF, UKF, QKF, and PF are defined in Table \ref{tab_paramObserver}. The initial conditions of the estimated states $\hat{\mathbf{x}}_{0}$ were established by sampling a normal distribution centered on the nominal initial conditions of the model, with a standard deviation of $0.01$. The covariance matrix $\mathbf{P}$ was initialized as a diagonal matrix using the observer’s initial state $\hat{\mathbf{x}}_{0}$. The values of $\mathbf{Q}$, $\mathbf{R}$, process noise $\bm{\omega}_{k}$, measurement noise $\bm{\upsilon}_{k}$, Kalman and Luenberger gains, as well as the design parameters of sigma points: $\alpha,\beta,\kappa$, the number of dimensions $\mathcal{M}$ of the QKF, and the number of particles $N$ of the PF are defined in Table \ref{tab_paramObserver} and are kept the same for each case study. 
\begin{table*}[ht]
\footnotesize
\renewcommand{\arraystretch}{1.5} 
\caption{Configuration parameters of the state observers under different cases. Common parameters: $\bm{\omega}_{k}  \sim \mathcal{N}(\bm{\omega}_{k};0, \mathbf{Q})$, $\bm{\upsilon}_{k} \sim  \mathcal{N}(\bm{\upsilon}_{k};0, \mathbf{R})$.}
\label{tab_paramObserver}
\centering
\begin{tabular}{|c|c|c|c|c|}
\hline
\textbf{Observer} & \textbf{Case 1} & \textbf{Case 2} & \textbf{Case 3} & \textbf{Special case} \\ 
\hline
\textbf{All} & 
\begin{tabular}{@{}c@{}}$n=5$\\$\mathbf{Q} = \textnormal{diag}(4.2\times10^{-8}) \in \mathbb{R}^{5\times1}$\\$\mathbf{R} = \textnormal{diag}(3.5\times10^{-8}) \in \mathbb{R}^{3\times1}$\end{tabular} &
\begin{tabular}{@{}c@{}}$n=10$\\$\mathbf{Q} = \textnormal{diag}(4.2\times10^{-6}) \in \mathbb{R}^{10\times1}$\\$\mathbf{R} = \textnormal{diag}(3.5\times10^{-6}) \in \mathbb{R}^{7\times1}$\end{tabular} &
\begin{tabular}{@{}c@{}}$n=15$\\$\mathbf{Q} = \textnormal{diag}(4.2\times10^{-6}) \in \mathbb{R}^{15\times1}$\\$\mathbf{R} = \textnormal{diag}(3.5\times10^{-6}) \in \mathbb{R}^{9\times1}$\end{tabular} &
\begin{tabular}{@{}c@{}}$n=3$\\$\mathbf{Q} = \textnormal{diag}(4.2\times10^{-8}) \in \mathbb{R}^{3\times1}$\\$\mathbf{R} = \textnormal{diag}(3.5\times10^{-8}) \in \mathbb{R}^{1\times1}$\end{tabular} \\
\hline
\textbf{ELO} & $\mathbf{L} = \mathbf{K}_{\infty} \in \mathbb{R}^{5\times3}$ & $\mathbf{L} = \mathbf{K}_{\infty} \in \mathbb{R}^{10\times7}$ & $\mathbf{L} = \mathbf{K}_{\infty} \in \mathbb{R}^{15\times9}$ & $\mathbf{L} = \mathbf{K}_{\infty} \in \mathbb{R}^{3\times1}$ \\ 
\hline
\textbf{EKF} & $\mathbf{K} \in \mathbb{R}^{5\times3}$ (Eq.~(\ref{eq8c})) & $\mathbf{K} \in \mathbb{R}^{10\times7}$ (Eq.~(\ref{eq8c})) & $\mathbf{K} \in \mathbb{R}^{15\times9}$ (Eq.~(\ref{eq8c})) & $\mathbf{K} \in \mathbb{R}^{3\times1}$ (Eq.~(\ref{eq8c})) \\ 
\hline
\textbf{UKF} & \multicolumn{4}{c|}{Parameters constant across all cases: $\alpha=1$, $\beta=2$, $\kappa=0$} \\ 
\hline
\textbf{QKF} & \multicolumn{4}{c|}{Parameters constant across all cases: $\mathcal{M}=3$, $\omega=[0.17,0.67,0.17]$} \\
\hline
\textbf{PF} & \multicolumn{4}{c|}{Particles number constant across all cases: $N=500$} \\
\hline
\end{tabular}
\end{table*}

\begin{figure}[h!]
	\centering
    {\includegraphics[width=0.95\columnwidth]{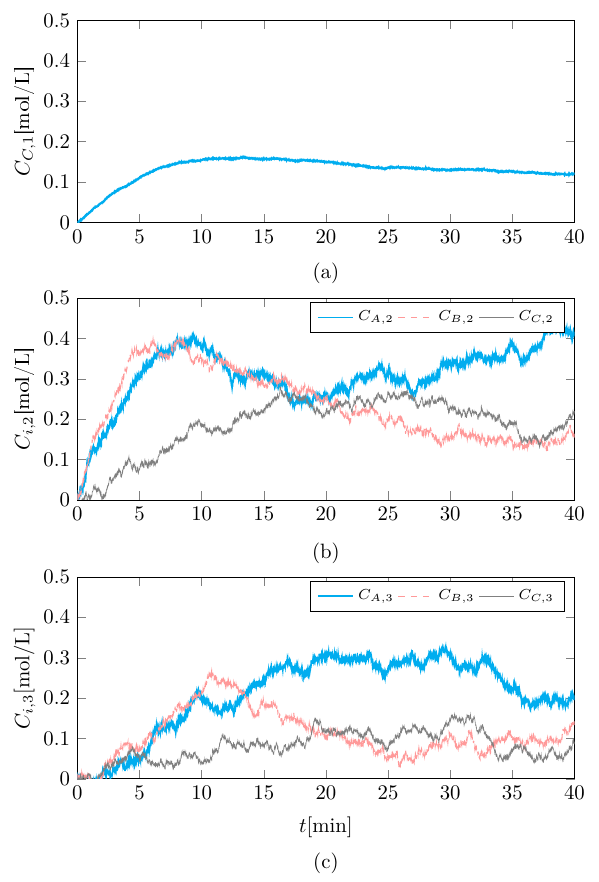}}
	\centering
	\caption{Available concentration measurements used for state estimation in each analyzed case. (a) Case 1: Measurement of $C_{C,1}$ (cyan). (b) Case 2: Measurements of $C_{A,2}$ (cyan), $C_{B,2}$ (dashed light red), and $C_{C,2}$ (grey). (c) Case 3: Measurements of $C_{A,3}$ (cyan), $C_{B,3}$ (dashed light red), and $C_{C,3}$ (grey).} 
\label{fig:mediciones}
\end{figure}

The time evolution of the measurable concentrations for Cases 1 to 3, obtained by simulating the CSTR system with the nonlinear model and parameter values from Table \ref{tabla3}, is shown in Figure \ref{fig:mediciones}. These correspond to the $\mathbf{y}$ vectors (Table \ref{tab_observability}) used for state estimation across all designed observers. Temperature measurements are shown in \ref{App_T}. 

Additionally, the sampling time used for all observers was set to $0.01$ [s], and the discretization method employed was the Euler approximation. This method was chosen due to its simplicity and low computational cost, making it suitable for real-time implementation, especially when using a sufficiently small sampling time.

\subsection{Performance criteria to evaluate the designed observers}

Two different reference signals can be used to assess the observers' performance. One approach is to compare the estimated state with the \textit{real state} of the mathematical model, which is assumed to be accurate. The other is to use noisy measurements that simulate real-world sensor data.

In this paper, both approaches are considered. In the first case, where the \textit{real state} of the model is available, the Mean Squared Error (MSE) is used as the performance metric. This metric quantifies the average squared difference between the estimated and true states across four test cases and is defined as:
\begin{equation}
\label{eq_MSE}
    \text{MSE} = \frac{1}{M} \sum_{i=1}^{M} (\mathbf{x}_{k,i} - \hat{\mathbf{x}}_{k,i})^2,
\end{equation}
where \( M \) is the total number of samples, \( \mathbf{x}_{k,i} \) is the true value of the \( i \)-th state, and \( \hat{\mathbf{x}}_{k,i} \) is the corresponding estimated value.

However, in practice, the complete state vector is not measured; this is the main issue leading to the need for state observers. Consequently, a more realistic evaluation for real-time applications relies on a second approach, which uses only the noisy output measurements to reference how well the observers estimate the states. This aligns better with practical scenarios where switching must be performed based on real-time measurements rather than full-state knowledge. To assess observer performance in this context, the following error-based metrics are proposed:
\begin{equation} 
\label{eq_L2} 
L_{2}^{(j)} = \frac{1}{n} \sum_{i=1}^{n} \sqrt{ \left| \mathbf{e}_{k,i}^{(j)} \right|^2 },
\end{equation}
\begin{equation} 
\label{eq_L8} 
L_{\infty}^{(j)} = \max\left( \left| \mathbf{e}_{k,i}^{(j)} \right| \right),
\end{equation}
where \( \mathbf{e}_{k,i}^{(j)} \) denotes the \( i \)-th component of the output estimation error vector defined in \eqref{eq_error}, and subindex $j$ runs through the six observers (SWO included). The metrics $L_2$ and $L_\infty$ differ from those used in the cost function \eqref{eq40}, enriching the evaluation by incorporating multiple perspectives on the accuracy of the estimation. The $L_2$ norm measures the average error magnitude, emphasizing overall consistency in estimation. In contrast, the $L_\infty$ norm captures the worst-case error, highlighting the robustness of the observer in avoiding large deviations. The joint use of these norms enables a more comprehensive and informative assessment of observer performance under realistic noise conditions.

\subsection{Monte Carlo Simulations}

The accuracy of a state observer in the CSTR system depends critically on precise model parameters. Yet, in practice, reaction kinetics, heat-transfer coefficients, and activation energies are subject to uncertainty due to disturbances and equipment aging. A Monte Carlo–based uncertainty analysis was conducted for Case 2 (two CSTRs in series) to assess the observer's robustness to such parameter variations.

In this study, 100 independent parameter perturbations were generated by sampling each component of the variation vector \(\Delta \boldsymbol{\varphi}\) (Eq. \eqref{eq:delta_phi}) from a uniform distribution. For each trial, the sampled perturbations were applied separately to Reactor 1 and Reactor 2 parameters, and the joint system, including the state observer, was simulated. The resulting distributions of estimation error, convergence rate, and steady-state accuracy were analyzed to quantify the observer’s performance under realistic parametric uncertainty.
\begin{equation}
\Delta \boldsymbol{\varphi} = \left[ k_0, \ E, \ UA, \ \Delta H \right]^{T} \sim \mathcal{U}(-1000,\ 1000),
\label{eq:delta_phi}
\end{equation}
where \(\Delta \boldsymbol{\varphi}\) denotes the vector of parameter perturbations and \(\mathcal{U}(-1000,1000)\) represents a uniform distribution over the specified bounds for each parameter.

\section{Results and discussion}
\label{sec4}

This section evaluates the proposed Switching Observer (SWO) strategy and compares its performance with individual observers (ELO, EKF, UKF, QKF, and PF) across different scenarios. The results are structured to analyze estimation accuracy, computational efficiency, robustness to parameter uncertainty, and practical feasibility for real-time applications.

\subsection{The need for a switching}
\begin{figure}[h]
	\centering   
\includegraphics[width=0.95\columnwidth]{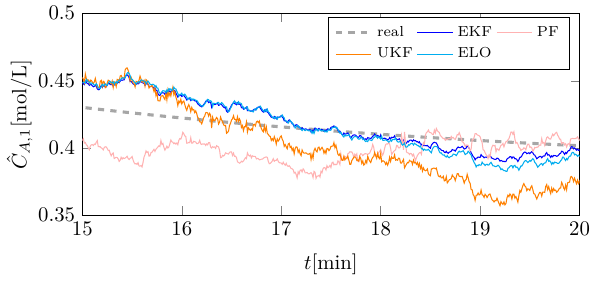}
	\centering
    \caption{Estimation of concentration $\hat{C}_{A,1}$ for Case 3 using a non-switched approach with five observers. True state (dashed grey), EKF (blue), PF (light red), UKF (orange), and ELO (cyan).} 
	\label{fig:Est_caso3_mot}
\end{figure}

Figure \ref{fig:Est_caso3_mot} shows the estimated concentration $\hat{C}_{A,1}$ in Case 3 when the five proposed observers operated independently, i.e., in a non-switched form. It is well known that each of these observers offers distinct advantages and limitations regarding the system's nonlinearity and measurement noise.

\begin{figure}[!t]
	\centering   
    {\includegraphics[width=\columnwidth]{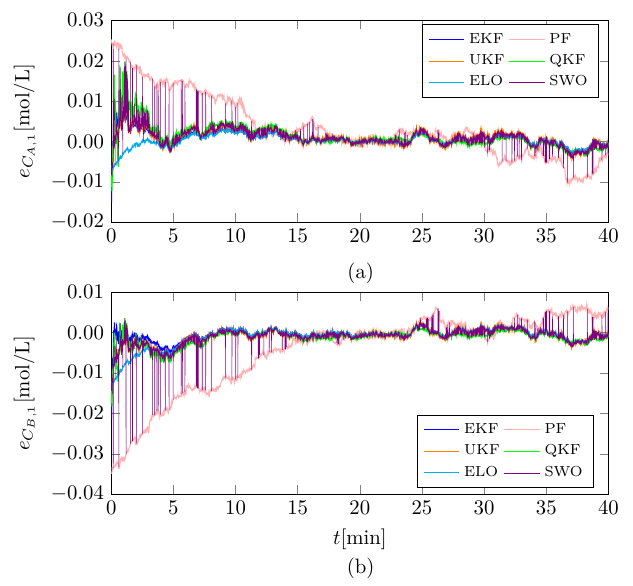}} 
	\centering
    \caption{Estimated concentrations in Case 1 using the proposed switched observer (SWO) versus non-switched observers. (a) Estimation error $e_{C_{A,1}}$ for $\hat{C}{A,1}$; (b) Estimation error $e{C_{B,1}}$ for $\hat{C}_{B,1}$. True values (dashed grey), EKF (blue), PF (light red), UKF (orange), QKF (green), ELO (cyan), and SWO (violet).} 
	\label{fig:Est_caso1_hat}
\end{figure}

Figure \ref{fig:Est_caso3_mot} shows that, although the estimates generated by the observers are generally similar, notable differences emerge upon closer examination. For instance, between the 15th and 16th minutes, the estimates deviate from the real value; then, during the interval between the 16th and 18th minutes, they converge again toward the real value. Finally, from the 18th minute onward, they diverge again. These discrepancies, although small, result in deviations that hinder consistent convergence to the real values throughout the system's operating range.

To address these limitations, we propose the switching observer described in Section \ref{sec31}, which dynamically selects the output of the observer with the best performance in real time, based on the estimation error criteria detailed in \eqref{eq40}. The purpose is then to leverage the individual strengths of each observer under varying operational conditions, resulting in an estimated signal that closely aligns with the system's true behavior.

\subsection{Testing the observers}

After running all simulations with the designed observers for Cases 1 to 3 and for the Special Case, the evolution of estimated states is plotted to analyze the results. The following is a summary of the plots and tables presented, which are used for further analysis.
\begin{figure}[!t]
	\centering   
    {\includegraphics[width=\columnwidth]{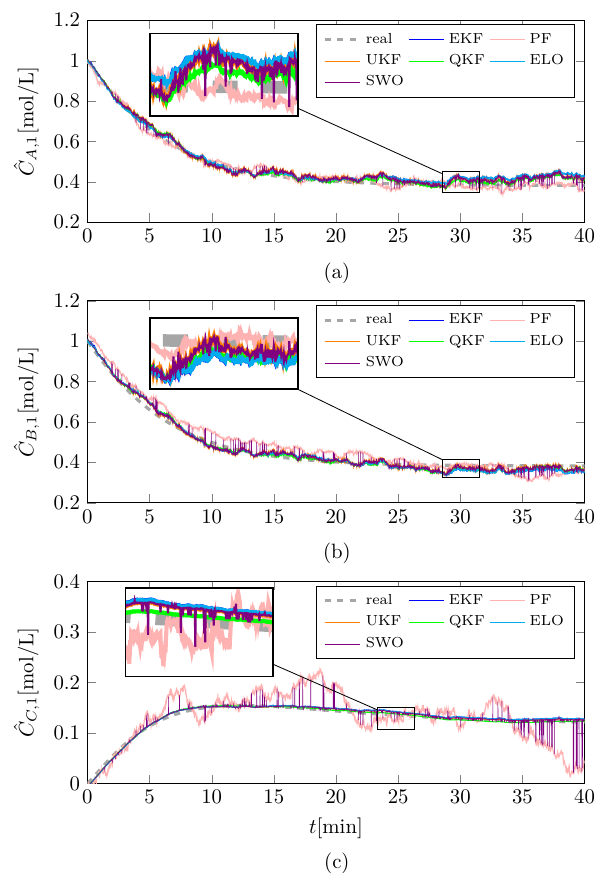}} 
	\centering
    \caption{Estimated concentrations in Case 2 using the proposed switched observer (SWO) compared to non-switched observers. (a) Concentration estimate of \( C_{A,1}\); (b) \(C_{B,1}\); (c) \(C_{C,1}\). EKF (dashed blue), PF (dashed grey), UKF (dashed orange), QKF (dashed green), ELO (dashed cyan), and SWO (violet).}   
	\label{fig:Est_caso2_hat}
\end{figure}
\begin{itemize}
    \item Figure \ref{fig:Est_caso1_hat} shows the evolution of the estimated concentrations $\hat{C}_{A,1}$ (a) and $\hat{C}_{B,1}$ (b) of the CSTR system for Case 1 (1 Reactor), which are estimated using the available measurements of $C_{C,1}$, $T_{1}$, and $T_{j,1}$.
    \item Figure \ref{fig:Est_caso2_hat} shows the resulting estimated states for Case 2, where the concentrations of Reactor 1 ($\hat{C}_{A,1}$ (a), $\hat{C}_{B,1}$ (b) and $\hat{C}_{C,1}$ (c)) are reconstructed using the concentrations of Reactor 2, along with the internal and jacket temperature measurements from both reactors.
    \item Figure \ref{fig:Est_caso3_hat} presents the results obtained for Case 3. Here, all concentrations produced in Reactors 1 ($\hat{C}_{A,1}$ (a), $\hat{C}_{B,1}$ (b) and $\hat{C}_{C,1}$) and Reactor 2 ($\hat{C}_{A,2}$ (d), $\hat{C}_{B,2}$ (e) and $\hat{C}_{C,2}$ (f))  are estimated from the concentrations in Reactor 3.
    \item Figure \ref{fig:Est_casoLin_hat} shows the evolution of the estimated states of the CSTR system for the Special Case, where a linear model of Reactor 1 is considered for design purposes. In this special case, only the measurements of $C_{C,1}$ are considered to obtain the concentration signals $\hat{C}_{A,1}$ (a) and $\hat{C}_{B,1}$ (b).
    \item Table \ref{tab:observer_times} shows the execution time corresponding to every observer used, representing the computational cost of each strategy. The computational performance of the proposed algorithm was assessed using MATLAB’s \textit{tic} and \textit{toc} functions, which measure the elapsed time between the start and end of code execution. This metric provides insight into the algorithm's computational efficiency.
    \item Tables \ref{tab:L2_all_cases} and \ref{tab:L8_all_cases} show the resulting values for performance indices $L_2$ \eqref{eq_L2} and $L_\infty$ \eqref{eq_L8} for all simulated cases, respectively.
    \item Table \ref{tab:MSE_all_cases} presents the resulting values for the MSE performance index \eqref{eq_MSE} for all cases. As mentioned above, this performance index is based on the differences between the estimated states for each case and the corresponding true states resulting from the simulated process, which serve as the \textit{ideal references} and are represented as dashed gray curves labeled ``real'' in Figures \ref{fig:Est_caso1_hat}, \ref{fig:Est_caso2_hat}, \ref{fig:Est_caso3_hat}, and \ref{fig:Est_casoLin_hat}. This visual representation allows for a direct comparison between the estimated and ideal system dynamics. The subtraction of each state vector $\mathbf{x}_{i}$ for $i=0,1,2,3$ in the MSE equation is valid because all observers are full-rank, ensuring they reconstruct estimated vectors of the same size as the corresponding state vectors. 
\end{itemize} 

\begin{figure*}[!t]
	\centering     
    {\includegraphics[width=2.1\columnwidth]{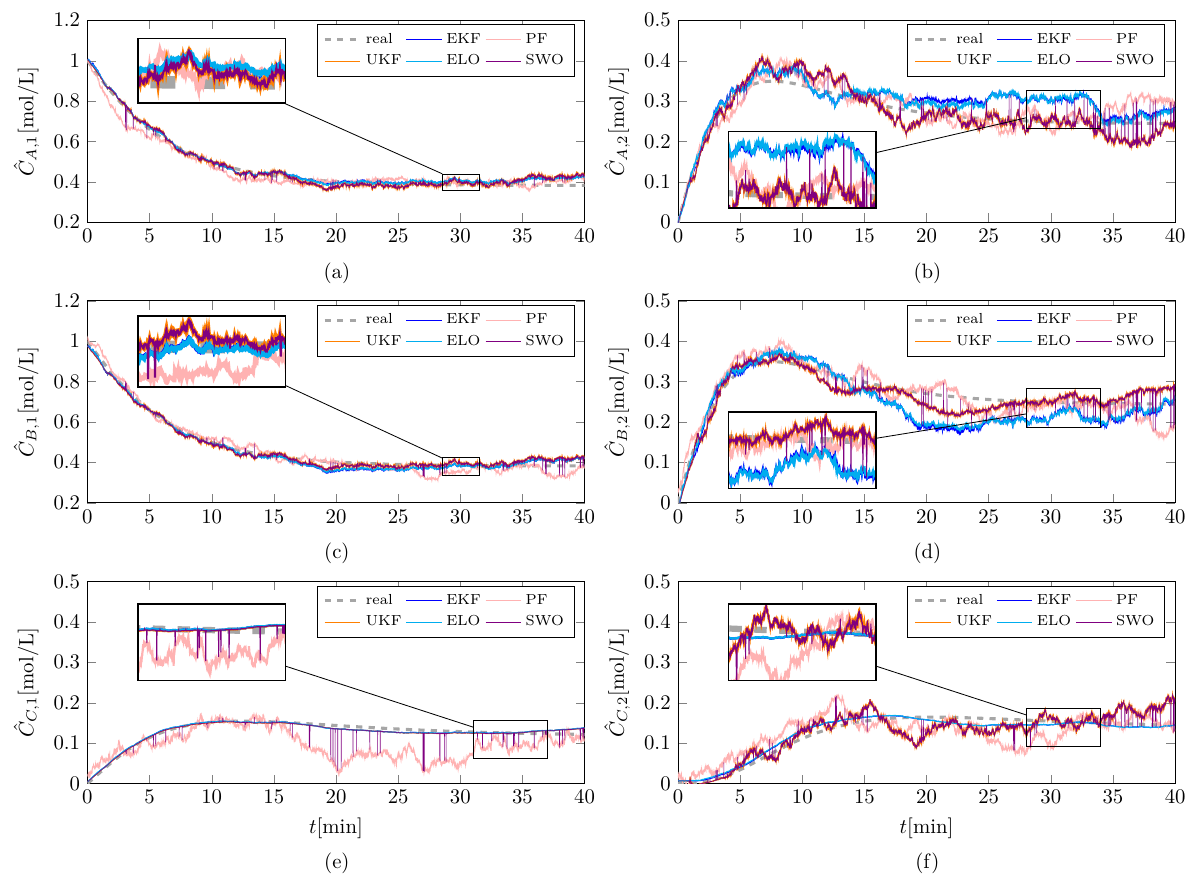}} 
	\centering
     \caption{Estimated concentrations in Case 3 using the proposed switched observer (SWO) compared to non-switched observers. (a) Concentration estimate of \( C_{A,1} \); (b) \(C_{B,1}\); (c) \(C_{C,1}\); (d) \(C_{A,2}\); (e) \(C_{B,2}\); (f) \(C_{C,2}\). EKF (dashed blue), PF (dashed grey), UKF (dashed orange), QKF (dashed green), ELO (dashed cyan), and SWO (violet).} 
	\label{fig:Est_caso3_hat}
\end{figure*}

\begin{figure}[!t]
	\centering   
    {\includegraphics[width=\columnwidth]{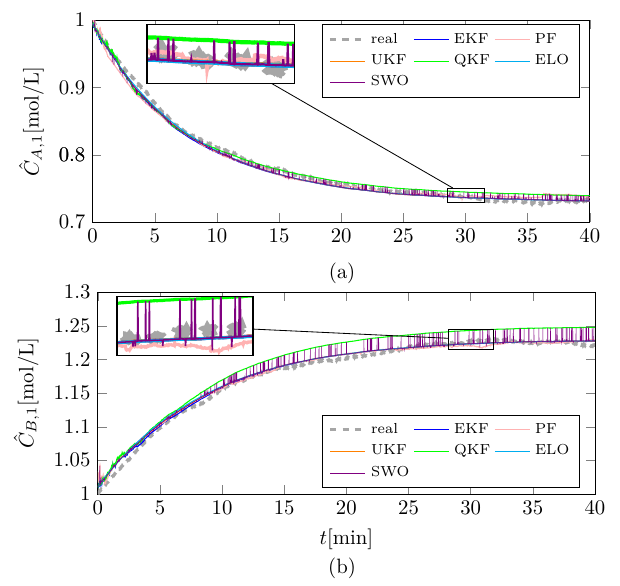}} 
	\centering
	\caption{Estimated concentrations in the Special Case using the proposed switched observer (SWO) compared to non-switched observers. (a) Concentration estimate of \(C_{A,1}\); (b) \(C_{B,1}\). True values (dashed grey), EKF (blue), PF (light red), UKF (orange), QKF (green), ELO (cyan), and SWO (violet).}  
	\label{fig:Est_casoLin_hat}
\end{figure}

\begin{table}[H]
\centering
\caption{Execution time for different observers in each case.}
\label{tab:observer_times}
\resizebox{\columnwidth}{!}{%
\begin{tabular}{|c|c|c|c|c|c|c|}
\hline
\textbf{Case} & \textbf{EKF [s]} & \textbf{PF [s]} & \textbf{UKF [s]} & \textbf{QKF [s]} & \textbf{ELO [s]} & \textbf{SW [s]} \\ \hline
1 & \cellcolor{gray!30} \textbf{1.055} & 5.018 & 1.218 & 2.838 & 1.118 & 7.062 \\ \hline
2 & 1.644 & 5.699 & 1.422 & 365.903 & \cellcolor{gray!30} \textbf{1.185} & 371.542 \\ \hline
3 & 1.182 & 6.074 & 1.572 & - & \cellcolor{gray!30} \textbf{1.156} & 8.673 \\ \hline
\end{tabular}%
}
\end{table}

\begin{table}[H]
\centering
\caption{L2 for different observers across all cases.}
\label{tab:L2_all_cases}
\resizebox{\columnwidth}{!}{%
\begin{tabular}{|c|c|c|c|c|c|c|}
\hline
 & \textbf{EKF} & \textbf{PF} & \textbf{UKF} & \textbf{QKF} & \textbf{ELO} & \textbf{SWO} \\ \hline
\multicolumn{7}{|c|}{\textbf{Case 1}} \\ \hline
$\hat{C}_{C,1}$ & $3.0\times 10^{-3}$ & $6.6\times 10^{-4}$ & $3.3\times 10^{-4}$ & $1.8\times 10^{-4}$ & $1.1\times 10^{-4}$ & \cellcolor{gray!30} $\mathbf{3.9\times 10^{-7}}$  \\ \hline

\multicolumn{7}{|c|}{\textbf{Case 2}} \\ \hline
$\hat{C}_{A,2}$ & $1.8\times 10^{-3}$ & $9.2\times 10^{-4}$ & $1.0\times 10^{-3}$ & $6.9\times 10^{-4}$ & $1.9\times 10^{-3}$ & \cellcolor{gray!30} $\mathbf{8.5\times 10^{-5}}$ \\ \hline

$\hat{C}_{B,2}$ & $2.0\times 10^{-3}$ & $8.6\times 10^{-4}$ & $9.2\times 10^{-4}$ & $1.0\times 10^{-3}$ & $7.6\times 10^{-4}$ & \cellcolor{gray!30} $\mathbf{1.0\times 10^{-5}}$ \\ \hline

$\hat{C}_{C,2}$ & $1.5\times 10^{-3}$ & $1.8\times 10^{-3}$ & $1.2\times 10^{-3}$ & $1.3\times 10^{-3}$ & $1.5\times 10^{-3}$ & \cellcolor{gray!30} $\mathbf{1.0 \times 10^{-4}}$ \\ \hline

\multicolumn{7}{|c|}{\textbf{Case 3}} \\ \hline
$\hat{C}_{A,3}$ & $2.8\times 10^{-3}$ & $1.4\times 10^{-3}$ & $1.3\times 10^{-3}$ & - & $1.2\times 10^{-3}$ & \cellcolor{gray!30} $\mathbf{7.2 \times 10^{-5}}$ \\ \hline

$\hat{C}_{B,3}$ & $2.5\times 10^{-3}$ & $1.2\times 10^{-3}$ & $1.7\times 10^{-3}$ & - & $1.4\times 10^{-3}$ & \cellcolor{gray!30} $\mathbf{6.8 \times 10^{-5}}$ \\ \hline

$\hat{C}_{C,3}$ & $2.1\times 10^{-3}$ & $1.0\times 10^{-3}$ & $1.4\times 10^{-3}$ & - & $1.6\times 10^{-3}$ & \cellcolor{gray!30} $\mathbf{8.5 \times 10^{-5}}$ \\ \hline

\multicolumn{7}{|c|}{\textbf{Special Case}} \\ \hline
$\hat{C}_{C,1}$ & $1.4\times 10^{-3}$ & $3.0\times 10^{-5}$ & $1.7\times 10^{-4}$ & $4.8\times 10^{-5}$ & $1.9\times 10^{-5}$ & \cellcolor{gray!30} $\mathbf{2.4\times 10^{-7}}$ \\ \hline
\end{tabular}%
}
\end{table}

\subsection{Analysis of the observers´ performance}

In the figure corresponding to Case 1 (Figure \ref{fig:Est_caso1_hat}), only the estimation errors are presented, in contrast to the Figures for Cases 2 and 3, which include the estimated states and their temporal evolution. This decision was made because the state estimations in Case 1 exhibited minimal visible variation, making it difficult to assess the observers' performance. In contrast, plotting the estimation errors provided a more precise and more informative visualization.

\begin{table}[H]
\centering
\caption{$L_\infty$ for different observers across all cases.}
\label{tab:L8_all_cases}
\resizebox{\columnwidth}{!}{%
\begin{tabular}{|c|c|c|c|c|c|c|}
\hline
 & \textbf{EKF} & \textbf{PF} & \textbf{UKF} & \textbf{QKF} & \textbf{ELO} & \textbf{SWO} \\ \hline
\multicolumn{7}{|c|}{\textbf{Case 1}} \\ \hline
$\hat{C}_{C,1}$ & $6.3\times 10^{-3}$ & $9.7\times 10^{-4}$ & $5.0\times 10^{-4}$ & $1.8\times 10^{-4}$ & $2.5\times 10^{-4}$ & \cellcolor{gray!30} $\mathbf{9.3\times 10^{-7}}$  \\ \hline

\multicolumn{7}{|c|}{\textbf{Case 2}} \\ \hline
$\hat{C}_{A,2}$ & $4.6\times 10^{-3}$ & $2.1\times 10^{-3}$ & $2.1\times 10^{-3}$ & $1.5\times 10^{-3}$ & $3.2\times 10^{-3}$ & \cellcolor{gray!30} $\mathbf{1.8\times 10^{-4}}$ \\ \hline

$\hat{C}_{B,2}$ & $4.8\times 10^{-3}$ & $1.8\times 10^{-3}$ & $2.0\times 10^{-3}$ & $2.5\times 10^{-3}$ & $2.0\times 10^{-3}$ & \cellcolor{gray!30} $\mathbf{2.4\times 10^{-4}}$ \\ \hline

$\hat{C}_{C,2}$ & $3.9\times 10^{-3}$ & $2.7\times 10^{-3}$ & $2.1\times 10^{-3}$ & $2.9\times 10^{-3}$ & $2.6\times 10^{-3}$ & \cellcolor{gray!30} $\mathbf{2.3 \times 10^{-4}}$ \\ \hline

\multicolumn{7}{|c|}{\textbf{Case 3}} \\ \hline
$\hat{C}_{A,3}$ & $6.6\times 10^{-3}$ & $2.5\times 10^{-3}$ & $3.5\times 10^{-3}$ & - & $2.7\times 10^{-3}$ & \cellcolor{gray!30} $\mathbf{2.1 \times 10^{-4}}$ \\ \hline

$\hat{C}_{B,3}$ & $5.2\times 10^{-3}$ & $3.5\times 10^{-3}$ & $2.8\times 10^{-3}$ & - & $3.0\times 10^{-3}$ & \cellcolor{gray!30} $\mathbf{1.3 \times 10^{-4}}$ \\ \hline

$\hat{C}_{C,3}$ & $6.6\times 10^{-3}$ & $1.4\times 10^{-3}$ & $2.1\times 10^{-3}$ & - & $3.0\times 10^{-3}$ & \cellcolor{gray!30} $\mathbf{2.1 \times 10^{-4}}$ \\ \hline

\multicolumn{7}{|c|}{\textbf{Special Case}} \\ \hline
$\hat{C}_{C,1}$ & $1.4\times 10^{-3}$ & $3.0\times 10^{-5}$ & $1.8\times 10^{-4}$ & $4.7\times 10^{-5}$ & $1.9\times 10^{-5}$ & \cellcolor{gray!30} $\mathbf{2.4\times 10^{-7}}$ \\ \hline
\end{tabular}%
}
\end{table}

\begin{table}[b]
\centering
\caption{MSE for different observers across all cases.}
\label{tab:MSE_all_cases}
\resizebox{\columnwidth}{!}{%
\begin{tabular}{|c|c|c|c|c|c|c|}
\hline
 & \textbf{EKF} & \textbf{PF} & \textbf{UKF} & \textbf{QKF} & \textbf{ELO} & \textbf{SWO} \\ \hline
\multicolumn{7}{|c|}{\textbf{Case 1}} \\ \hline
$\hat{C}_{A,1}$ & $3.1\times 10^{-6}$ & $7.3\times 10^{-5}$ & $4.2\times 10^{-6}$ & \cellcolor{gray!30} $\mathbf{1.1\times 10^{-6}}$ & $2.2 \times 10^{-6}$ & $7.5\times 10^{-6}$ \\ \hline
$\hat{C}_{B,1}$ & \cellcolor{gray!30} $\mathbf{2.2\times 10^{-6}}$ & $1.2\times 10^{-4}$ & $4.2\times 10^{-6}$ & $5.2\times 10^{-6}$ & $6.8\times 10^{-6}$ & $7.4\times 10^{-6}$ \\ \hline
\multicolumn{7}{|c|}{\textbf{Case 2}} \\ \hline
$\hat{C}_{A,1}$ & $7.7\times 10^{-4}$ & \cellcolor{gray!30} $\mathbf{2.7\times 10^{-4}}$ & $6.3\times 10^{-4}$ & $3.6\times 10^{-4}$ & $7.5\times 10^{-4}$ & $6.0\times 10^{-4}$ \\ \hline
$\hat{C}_{B,1}$ & $4.4\times 10^{-4}$ & $1.8\times 10^{-3}$ & \cellcolor{gray!30} $\mathbf{3.1\times 10^{-4}}$ & $3.4\times 10^{-4}$ & $4.6\times 10^{-4}$ & $3.7\times 10^{-4}$ \\ \hline
$\hat{C}_{C,1}$ & $1.9\times 10^{-5}$ & $1.2\times 10^{-3}$ & $1.3\times 10^{-5}$ & \cellcolor{gray!30} $\mathbf{7.7\times 10^{-6}}$ & $1.8\times 10^{-5}$ & $5.1\times 10^{-5}$ \\ \hline
\multicolumn{7}{|c|}{\textbf{Case 3}} \\ \hline
$\hat{C}_{A,1}$ & \cellcolor{gray!30} $\mathbf{2.9\times 10^{-4}}$ & $8.9\times 10^{-4}$ & $4.3\times 10^{-4}$ & - & $3.0\times 10^{-4}$ & $4.3\times 10^{-4}$ \\ \hline
$\hat{C}_{B,1}$ & $4.3\times 10^{-4}$ & $6.9\times 10^{-4}$ & \cellcolor{gray!30} $\mathbf{3.6\times 10^{-4}}$ & - & $4.0\times 10^{-4}$ & \cellcolor{gray!30} $\mathbf{3.6\times 10^{-4}}$ \\ \hline
$\hat{C}_{C,1}$ & \cellcolor{gray!30} $\mathbf{3.0\times 10^{-5}}$ & $2.2\times 10^{-3}$ & \cellcolor{gray!30} $\mathbf{3.0\times 10^{-5}}$ & - & $3.2\times 10^{-5}$ & $6.7\times 10^{-5}$ \\ \hline
$\hat{C}_{A,2}$ & $1.2\times 10^{-3}$ & \cellcolor{gray!30} $\mathbf{7.9\times 10^{-4}}$ & $6.9\times 10^{-4}$ & - & $1.2\times 10^{-3}$ & $7.0\times 10^{-4}$ \\ \hline
$\hat{C}_{B,2}$ & $1.7\times 10^{-3}$ & $7.0\times 10^{-4}$ & \cellcolor{gray!30} $\mathbf{4.9\times 10^{-4}}$ & - & $1.6\times 10^{-3}$ & $5.1\times 10^{-4}$ \\ \hline
$\hat{C}_{C,2}$ & \cellcolor{gray!30} $\mathbf{1.1\times 10^{-4}}$ & $1.3\times 10^{-3}$ & $6.9\times 10^{-4}$ & - & $1.2\times 10^{-4}$ & $6.8\times 10^{-4}$ \\ \hline
\multicolumn{7}{|c|}{\textbf{Special Case}} \\ \hline
$\hat{C}_{A,1}$ & $3.4\times 10^{-5}$ & $5.5\times 10^{-5}$ & $2.4\times 10^{-5}$ & $4.7\times 10^{-5}$ & \cellcolor{gray!30} $\mathbf{1.9\times 10^{-5}}$ & $2.6\times 10^{-5}$ \\ \hline
$\hat{C}_{B,1}$ & \cellcolor{gray!30} $\mathbf{2.3\times 10^{-5}}$ & $3.8\times 10^{-5}$ & $2.9\times 10^{-5}$ & $3.4\times 10^{-4}$ & $3.5\times 10^{-5}$ & $4.3\times 10^{-5}$ \\ \hline
\end{tabular}%
}
\end{table}

Figures \ref{fig:Est_caso1_hat} (a)–(b) present the estimation errors of 
\( C_{A,1} \) and $C_{B,1}$ with respect to their true values for Case 1. These plots reveal that all estimation errors converge toward zero over time. The SWO error exhibits fluctuations at each time point, reflecting the dynamic selection of the optimal observer as defined by the cost function in Equation \eqref{eq40}. Overall, the SWO achieves superior performance in terms of the  $L_{2}$ and $L_\infty$ norms (see Tables \ref{tab:L2_all_cases} and \ref{tab:L8_all_cases}). However, with respect to the mean squared error (MSE), the SWO does not consistently achieve the best performance (see Table \ref{tab:MSE_all_cases}). Specifically, the QKF yields the lowest MSE for the estimation of $\hat{C}_{A,1}$, while the EKF provides the most accurate estimate for $\hat{C}_{B,1}$. The remaining observers (PF, UKF, and ELO) show comparable behavior. It is important to note that no single observer consistently achieves the lowest MSE across all cases, and the differences between their best MSE values and that of the SWO are neither significant nor substantial. In contrast, in every case, the SWO consistently outperforms all individual observers in terms of both $L_{2}$ and $L_\infty$ norms. These findings highlight the effectiveness of the switching strategy in minimizing overall estimation error.

In Case 2, as shown in Figure \ref{fig:Est_caso2_hat}, the estimates  $\hat{C}_{A,1}$ and $\hat{C}_{B,1}$ for all the observers follow the same dynamics as the real concentrations that serve as reference (grey curve). However, in the case of $\hat{C}_{C,1}$, there is evidence of large estimation errors for certain intervals where the estimated signals are far from the real value, especially for the PF observer. This shows the complexity of the estimation when there are two reactors in series. Tables \ref{tab:L2_all_cases} and \ref{tab:L8_all_cases} indicate that for Case 2, the best indices $L_{2}$ and $L_\infty$ correspond to the SWO. However, in Table \ref{tab:MSE_all_cases}, which presents the mean squared error (MSE), the SWO does not have the lowest error. There is variability in the best observer: for the estimation of $\hat{C}_{A,1}$, the best is the PF; for $\hat{C}_{B,1}$, it is the UKF; and for $\hat{C}_{C,1}$, the QKF prevails. It is important to note that the QKF takes the longest computation time during the simulation at $365.9$ [s] (as shown in Table \ref{tab:observer_times}), but it provides the best estimate of the $\hat{C}_{C,1}$ concentration.

Case 3 represents the most complex scenario analyzed in this study, which can be observed from Figure \ref{fig:Est_caso3_hat}. The state of Reactors 1 and 2 is estimated based on the concentrations in Reactor 3 and the internal and jacket temperatures of all reactors (as shown in Table \ref{tab_observability}).
Generally, the estimates for $\hat{C}_{A,1}$, $\hat{C}_{B,1}$, and $\hat{C}_{C,1}$ (Figures \ref{fig:Est_caso3_hat} (a), (c), and (e)) behave similarly to those observed in Case 2; however, $\hat{C}_{C,1}$ exhibits greater fluctuations. As illustrated in Figures \ref{fig:Est_caso3_hat} (b), (d), and (f), the concentrations in Reactor 2 generally track the actual values. However, the behavior is less conservative than Reactor 1 due to the increased fluctuations. Notably, the SWO observer demonstrates the lowest $L_{2}$ and $L_\infty$ values in this case (refer to Table \ref{tab:L2_all_cases} and \ref{tab:L8_all_cases}), highlighting its ability to recover from the noise and complexity of the scenario. The EKF and UKF achieved the best mean square error (MSE) values (see Table \ref{tab:MSE_all_cases}), while the PF emerged as the only optimal approach for estimating $\hat{C}_{A,2}$. Conversely, the ELO observer exhibited poor performance in this case. However, for Case 3, the QKF observer failed to function effectively. This is mainly due to the computational burden. Since the execution time for 10 states in Case 2 was already significantly high, the increase to 15 states in Case 3 made the simulation time grow disproportionately, rendering the QKF impractical for this configuration. The high-dimensional state space dramatically amplifies the computational demands of the quadrature-based approach, leading to unacceptable delays and making it unsuitable for real-time or large-scale implementations in this scenario.

For the special case, it can be observed in Figure \ref{fig:Est_casoLin_hat} that the results closely resemble those obtained in Case 1, which leads us to conclude that the linearization performed is acceptable and that the CSTR can be represented in both modalities. In this case, the switching strategy remained the same as in the nonlinear cases. Still, the design of the individual observers was derived from the linear versions of the proposed observers. The ELO  performs best in the Special Case for $\hat{C}_{A,1}$, suggesting that it may be helpful for linearized systems.

In terms of how often observers were used in the switched strategy, we can mention that the UKF was the most used observer (86.62\% of the time), indicating its robustness and reliability. The QKF, although it was the second most used (8.39\% of the time), was the slowest, with very long execution times (e.g., $365.9$ [s] in Case 2), even making its use infeasible in Case 3 (see Table \ref{tab:observer_times} for detailed execution times in each case).

The execution time Table \ref{tab:observer_times} reveals significant differences in the computational performance of the evaluated observers. Overall, the EKF stands out as one of the most efficient, being the fastest in Case 1 and the second fastest in Case 3. However, the ELO observer achieved the shortest execution time in Cases 2 and 3 and ranked second fastest in Case 1, making it the most balanced option regarding speed, but not in terms of reliability. The PF and QKF observers also present efficiency limitations, ranking second slowest in Cases 1 and 2, respectively. These results suggest that EKF and UKF are preferable choices for real-time applications or scenarios with limited computational resources. Although the computational time for the SWO was higher than that of individual filters (e.g., $371.5$ [s] in Case 2), this value reflects the combined execution of all filters. It is worth noting that the excessive execution time is primarily due to the QKF, which is not well-suited for high-dimensional systems such as in Cases 2 and 3. Therefore, the total execution time of the SWO should be considered comparable to that of other observers like the UKF and PF in such scenarios.

A key finding of this study is that the sigma-point approach outperforms other alternatives to the EKF’s local linearization, such as particle-based methods and quadrature-based transformations. In particular, the Unscented Kalman Filter (UKF) achieves high estimation accuracy with relatively low computational overhead, making it an efficient choice for nonlinear state estimation.
\begin{figure*}[ht]
    \centering
    \includegraphics[width=1\textwidth]{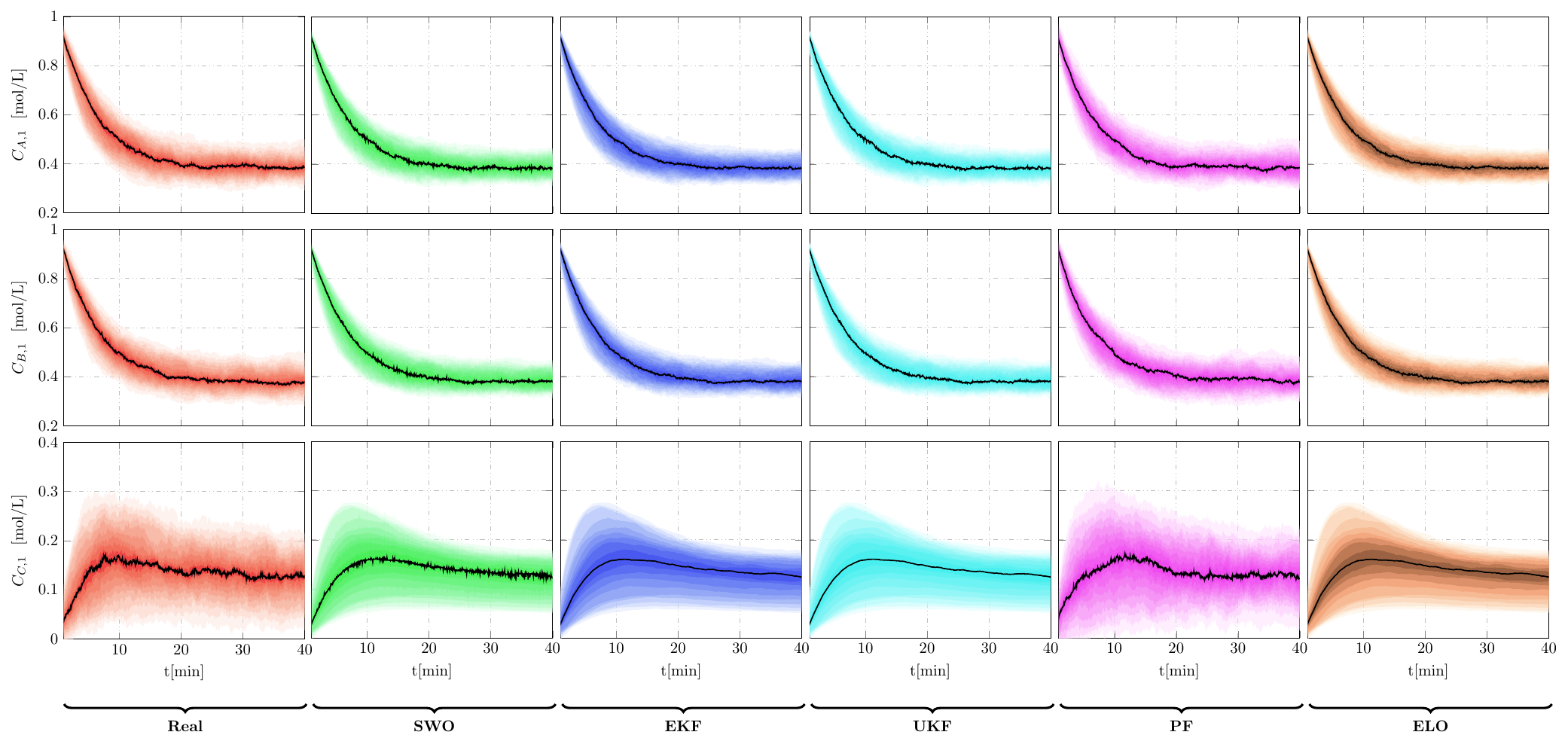}
    \caption{Results of the sensitivity simulations for all the observers in Case 2. (a,c,g) - \( C_{A,1} \) concentration estimation; (b,e,h) - $C_{B,1}$ concentration estimation; (c,f,i) - $C_{C,1}$ estimation. Real states (red), SWO (green), EKF (blue), UKF (cyan), PF (pink), and ELO (orange).}
    \label{fig:MontecarloEstado}
\end{figure*}

\subsection{Analysis of the parametric variation}

A Monte Carlo study was conducted to evaluate the robustness of the SWO under parametric uncertainty. Figure \ref{fig:MontecarloEstado} illustrates the behavior of the CSTR model and the observers under these parametric variabilities. The black curve at the center of the shaded area represents the average trend of each plot. In Figure \ref{fig:MontecarloEstado}, the first column represents the real values, shaded in red, followed in this order by the SWO (green), EKF (blue), UKF (cyan), PF (purple), and ELO (orange) estimates. The QKF could not be plotted in these tests, as it struggled to handle parametric variations effectively, highlighting the robustness of the other observers in this context.

From Figure \ref{fig:MontecarloEstado}, it can be observed that the estimates of $\hat{C}_{A,1},\hat{C}_{B,1}$ and $\hat{C}_{C,1}$ obtained with the switching observer are robust under parametric uncertainty with small variability in estimated states and rapid recovery after parameter perturbations.  
The remaining observers, EKF, UKF, and ELO, also show robustness against parametric uncertainty. Observer PF shows the greatest dispersion, implying that it is less robust in this test.

The comprehensive evaluation demonstrates that the proposed Switching Observer framework offers advantages in nonlinear, partially observable chemical processes by dynamically leveraging the strengths of different observers. The SWO maintains robustness under uncertainty and remains computationally feasible for real-time applications. In practical terms, the SWO could enable better soft sensor designs, improve fault detection capabilities, and support model-predictive control strategies in industrial chemical reactors with limited instrumentation.

\section{Conclusions}
\label{sec5}

This work has presented a novel switching observer framework for enhanced state estimation in nonlinear chemical reactor systems, specifically focusing on multi-stage CSTRs operating under partial observability. The proposed approach integrates five heterogeneous observer structures into a unified architecture: Extended Luenberger Observer, Extended Kalman Filter, Unscented Kalman Filter, Quadrature Kalman Filter, and Particle Filter. A composite cost function, combining the $L_1$ norm and Kullback–Leibler divergence of the output estimation error, governs the real-time selection of the most accurate observer at each sampling instant.

Simulation studies were conducted for four reactor configurations, ranging from a linearized single-reactor model to a nonlinear three-reactor network. The results demonstrate that the switching observer outperforms individual estimators regarding $L_2$ error, maximum error ($L_{\infty}$). The UKF emerged as the most frequently selected observer, but the switching strategy dynamically incorporated other estimators in transient or uncertain conditions, illustrating its adaptability.
The computational cost of the proposed framework remains compatible with typical sampling rates applied in chemical process control. Although the execution time increases due to the parallel operation of multiple observers, the switching strategy benefits from the modular nature of its design, allowing scalability and potential for hardware optimization. Moreover, Monte Carlo simulations confirmed the robustness of the switching strategy against parameter uncertainties, underscoring its applicability in real-world industrial scenarios.

Overall, this switching observer strategy offers a flexible and reliable alternative for state estimation in complex chemical systems with limited instrumentation. It can be a foundational layer for soft sensing, fault detection, and advanced control strategies in nonlinear and distributed process environments. Future work may consider experimental validation in pilot-scale setups, development of reduced-order observer banks for faster implementation, and integration with control frameworks such as model predictive control (MPC) under real-time constraints.

\section{Acknowledgment}
This work was partly supported by ANID-FONDECYT under grant numbers 3240317, 3240181, 3230398, and 1220168; and the ANID-Basal Project Grant AFB240002 (AC3E).

\bibliographystyle{elsarticle-num}
\bibliography{biblioteca_SW}

\begin{thebibliography}{10}
\expandafter\ifx\csname url\endcsname\relax
  \def\url#1{\texttt{#1}}\fi
\expandafter\ifx\csname urlprefix\endcsname\relax\def\urlprefix{URL }\fi
\expandafter\ifx\csname href\endcsname\relax
  \def\href#1#2{#2} \def\path#1{#1}\fi

\bibitem{kamen2012introduction}
E.~W. Kamen, J.~K. Su, Introduction to optimal estimation, Springer Science \&
  Business Media, 2012.

\bibitem{anderson2005optimal}
B.~D. Anderson, J.~B. Moore, Optimal filtering, Courier Corporation, 2005.

\bibitem{cortes_design_2021}
S.~Cortes, L.~E. Cortes, E.~Sierra~Vanegas, Design of a {Kalman} {Filter} and
  {Three} {Observers} in a {CSTR} for the {Estimation} of {Concentration} and
  {Temperature} in {Jacket}., International Journal on Advanced Science,
  Engineering and Information Technology 11~(4) (2021) 1405.
\newblock \href {https://doi.org/10.18517/ijaseit.11.4.13722}
  {\path{doi:10.18517/ijaseit.11.4.13722}}.

\bibitem{grewal2014kalman}
M.~S. Grewal, A.~P. Andrews, Kalman filtering: Theory and Practice with MATLAB,
  John Wiley \& Sons, 2014.

\bibitem{julier1997new}
S.~J. Julier, J.~K. Uhlmann, New extension of the kalman filter to nonlinear
  systems, in: Signal processing, sensor fusion, and target recognition VI,
  Vol. 3068, Spie, 1997, pp. 182--193.
\newblock \href {https://doi.org/https://doi.org/10.1117/12.280797}
  {\path{doi:https://doi.org/10.1117/12.280797}}.

\bibitem{farsi_state_2019}
M.~Farsi, M.~Dehghan~Manshadi, A {State} {Estimation} {Method} {Based} on
  {Integration} of {Linear} and {Extended} {Kalman} {Filters}, Chemical Product
  and Process Modeling 14~(4) (Dec. 2019).
\newblock \href {https://doi.org/10.1515/cppm-2018-0051}
  {\path{doi:10.1515/cppm-2018-0051}}.

\bibitem{chen_applying_2017}
J.~Chen, S.~Zhang, B.~Tong, Applying a cubature {Kalman} filter for the
  nonlinear state estimations of a continuous stirred tank reactor, in: 2017
  29th {Chinese} {Control} {And} {Decision} {Conference} ({CCDC}), IEEE,
  Chongqing, China, 2017, pp. 6343--6348.
\newblock \href {https://doi.org/10.1109/CCDC.2017.7978313}
  {\path{doi:10.1109/CCDC.2017.7978313}}.

\bibitem{arasaratnam2007discrete}
I.~Arasaratnam, S.~Haykin, R.~J. Elliott, Discrete-time nonlinear filtering
  algorithms using gauss--hermite quadrature, Proceedings of the IEEE 95~(5)
  (2007) 953--977.
\newblock \href {https://doi.org/10.1109/JPROC.2007.894705}
  {\path{doi:10.1109/JPROC.2007.894705}}.

\bibitem{gordon1993novel}
N.~J. Gordon, D.~J. Salmond, A.~F. Smith, Novel approach to
  nonlinear/non-gaussian bayesian state estimation, in: IEE proceedings F
  (radar and signal processing), Vol. 140, IET, 1993, pp. 107--113.
\newblock \href {https://doi.org/https://doi.org/10.1049/ip-f-2.1993.0015}
  {\path{doi:https://doi.org/10.1049/ip-f-2.1993.0015}}.

\bibitem{Zeitz1987}
M.~Zeitz, The extended luenberger observer for nonlinear systems, Systems \&
  Control Letters 9~(2) (1987) 149--156.
\newblock \href {https://doi.org/https://doi.org/10.1016/0167-6911(87)90021-1}
  {\path{doi:https://doi.org/10.1016/0167-6911(87)90021-1}}.

\bibitem{sanchez_torres_robust_2016}
J.~D. Sánchez~Torres, H.~A. Botero, E.~Jiménez, O.~Jaramillo, A.~G.
  Loukianov, A {Robust} {Extended} {State} {Observer} for the {Estimation} of
  {Concentration} and {Kinetics} in a {CSTR}, International Journal of Chemical
  Reactor Engineering 14~(1) (2016) 481--490.
\newblock \href {https://doi.org/10.1515/ijcre-2015-0149}
  {\path{doi:10.1515/ijcre-2015-0149}}.

\bibitem{medjebouri_extended_2023}
A.~Medjebouri, {Extended} {state} {observer} {nased} {robust} {feedback}
  {linearization} {control} {applied} {to} {an} {industrial} {CSTR}, Journal of
  Automation, Mobile Robotics and Intelligent Systems 17 (2023).
\newblock \href {https://doi.org/https://doi.org/10.14313/jamris/4-2023/32}
  {\path{doi:https://doi.org/10.14313/jamris/4-2023/32}}.

\bibitem{ling_state_2016}
C.~Ling, C.~Kravaris, State {Observer} {Design} for {Monitoring} the {Degree}
  of {Polymerization} in a {Series} of {Melt} {Polycondensation} {Reactors},
  Processes 4~(1) (2016) 4.
\newblock \href {https://doi.org/10.3390/pr4010004}
  {\path{doi:10.3390/pr4010004}}.

\bibitem{kumar_system_2024}
M.~Kumar, U.~Mehta, G.~Cirrincione, System identification of a nonlinear
  continuously stirred tank reactor using fractional neural network, South
  African Journal of Chemical Engineering 50 (2024) 299--310.
\newblock \href {https://doi.org/10.1016/j.sajce.2024.09.005}
  {\path{doi:10.1016/j.sajce.2024.09.005}}.

\bibitem{li_neural-network_2021}
S.~Li, C.~K. Ahn, J.~Guo, Z.~Xiang, Neural-{Network} {Approximation}-{Based}
  {Adaptive} {Periodic} {Event}-{Triggered} {Output}-{Feedback} {Control} of
  {Switched} {Nonlinear} {Systems}, IEEE Transactions on Cybernetics 51~(8)
  (2021) 4011--4020.
\newblock \href {https://doi.org/10.1109/TCYB.2020.3022270}
  {\path{doi:10.1109/TCYB.2020.3022270}}.

\bibitem{Bzioui_adaptive_2021}
S.~Bzioui, R.~Channa, An {Adaptive} {Observer} {Design} for {Nonlinear}
  {Systems} {Affected} by {Unknown} {Disturbance} with {Simultaneous}
  {Actuator} and {Sensor} {Faults}. {Application} to a {CSTR}, Biointerface
  Research in Applied Chemistry 12~(4) (2021) 4847--4856.
\newblock \href {https://doi.org/10.33263/BRIAC124.48474856}
  {\path{doi:10.33263/BRIAC124.48474856}}.

\bibitem{zhao_reaction_2016}
Z.~Zhao, J.~M. Wassick, J.~Ferrio, B.~E. Ydstie, Reaction variants and
  invariants based observer and controller design for {CSTRs}**{The} authors
  acknowledge {The} {Dow} {Chemical} {Company} for providing financial support
  for this work., IFAC-PapersOnLine 49~(7) (2016) 1091--1096.
\newblock \href {https://doi.org/10.1016/j.ifacol.2016.07.348}
  {\path{doi:10.1016/j.ifacol.2016.07.348}}.

\bibitem{subramanian_adaptive_2015}
S.~Subramanian, S.~Lucia, S.~Engel, Adaptive {Multi}-stage {Output} {Feedback}
  {NMPC} using the {Extended} {Kalman} {Filter} for time varying uncertainties
  applied to a {CSTR}, IFAC-PapersOnLine 48~(23) (2015) 242--247.
\newblock \href {https://doi.org/10.1016/j.ifacol.2015.11.290}
  {\path{doi:10.1016/j.ifacol.2015.11.290}}.

\bibitem{romero-bustamante_robust_2017}
J.~A. Romero-Bustamante, J.~G. Moguel-Castañeda, H.~Puebla,
  E.~Hernandez-Martinez, Robust {Cascade} {Control} for {Chemical} {Reactors}:
  {An} {Approach} based on {Modelling} {Error} {Compensation}, International
  Journal of Chemical Reactor Engineering 15~(6) (Dec. 2017).
\newblock \href {https://doi.org/10.1515/ijcre-2017-0082}
  {\path{doi:10.1515/ijcre-2017-0082}}.

\bibitem{ballesteros-moncada_fuzzy_2015}
H.~Ballesteros-Moncada, E.~J. Herrera-López, J.~Anzurez-Marín, Fuzzy
  model-based observers for fault detection in {CSTR}, ISA Transactions 59
  (2015) 325--333.
\newblock \href {https://doi.org/10.1016/j.isatra.2015.10.006}
  {\path{doi:10.1016/j.isatra.2015.10.006}}.

\bibitem{zarei_robust_2014}
J.~Zarei, E.~Shokri, Robust sensor fault detection based on nonlinear unknown
  input observer, Measurement 48 (2014) 355--367.
\newblock \href {https://doi.org/10.1016/j.measurement.2013.11.015}
  {\path{doi:10.1016/j.measurement.2013.11.015}}.

\bibitem{ahrens_high-gain_2009}
J.~H. Ahrens, H.~K. Khalil, High-gain observers in the presence of measurement
  noise: {A} switched-gain approach, Automatica 45~(4) (2009) 936--943.
\newblock \href {https://doi.org/10.1016/j.automatica.2008.11.012}
  {\path{doi:10.1016/j.automatica.2008.11.012}}.

\bibitem{barut_switching_2007}
M.~Barut, S.~Bogosyan, M.~Gokasan, Switching {EKF} technique for rotor and
  stator resistance estimation in speed sensorless control of {IMs}, Energy
  Conversion and Management 48~(12) (2007) 3120--3134.
\newblock \href {https://doi.org/10.1016/j.enconman.2007.04.026}
  {\path{doi:10.1016/j.enconman.2007.04.026}}.

\bibitem{Aguila_etal_2025}
M.~Fernández-Jorquera, M.~Zepeda-Rabanal, N.~Aguila-Camacho,
  L.~Bárzaga-Martell, Design, tuning, and experimental validation of switched
  fractional-order pid controllers for an inverted pendulum system, Fractal and
  Fractional 9~(4) (2025).
\newblock \href {https://doi.org/10.3390/fractalfract9040234}
  {\path{doi:10.3390/fractalfract9040234}}.

\bibitem{Afghoul2017}
H.~Afghoul, F.~Krim, A.~Beddar, M.~Houabes, Switched fractional order
  controller for grid connected wind energy conversion system, in: 2017 5th
  International Conference on Electrical Engineering - Boumerdes (ICEE-B),
  2017, pp. 1--5.
\newblock \href {https://doi.org/10.1109/ICEE-B.2017.8191970}
  {\path{doi:10.1109/ICEE-B.2017.8191970}}.

\bibitem{Aguila_Gallegos_2024}
N.~Aguila-Camacho, J.~A. Gallegos, Error-based switched fractional order model
  reference adaptive control for mimo linear time invariant systems, Fractal
  and Fractional 8~(2) (FEB 2024).
\newblock \href {https://doi.org/10.3390/fractalfract8020109}
  {\path{doi:10.3390/fractalfract8020109}}.

\bibitem{koumboulis_switching_2012}
F.~N. Koumboulis, D.~G. Fragkoulis, A switching observer design scheme for a
  double effect evaporator, in: 2012 {IEEE} {International} {Conference} on
  {Industrial} {Technology}, IEEE, Athens, 2012, pp. 650--655.
\newblock \href {https://doi.org/10.1109/icit.2012.6210012}
  {\path{doi:10.1109/icit.2012.6210012}}.

\bibitem{koumboulis_switching_2017}
F.~N. Koumboulis, D.~G. Fragkoulis, Switching design for the observation of the
  biomass in alcoholic fermentation processes, in: 2017 {XXVI} {International}
  {Conference} on {Information}, {Communication} and {Automation}
  {Technologies} ({ICAT}), IEEE, Sarajevo, 2017, pp. 1--6.
\newblock \href {https://doi.org/10.1109/icat.2017.8171621}
  {\path{doi:10.1109/icat.2017.8171621}}.

\bibitem{LiuSwitching}
Y.~Liu, Switching observer design for uncertain nonlinear systems, IEEE
  Transactions on Automatic Control 42~(12) (1997) 1699--1703.
\newblock \href {https://doi.org/10.1109/9.650020}
  {\path{doi:10.1109/9.650020}}.

\bibitem{kazantzis_nonlinear_2000}
N.~Kazantzis, C.~Kravaris, R.~A. Wright, Nonlinear {Observer} {Design} for
  {Process} {Monitoring}, Industrial \& Engineering Chemistry Research 39~(2)
  (2000) 408--419, publisher: American Chemical Society (ACS).
\newblock \href {https://doi.org/10.1021/ie990321n}
  {\path{doi:10.1021/ie990321n}}.

\bibitem{sainz-garcia_adaptive_2022}
S.~H. Sainz-García, G.~López~López, V.~M. Alvarado, J.~Y. Rumbo~Morales,
  E.~Sarmiento-Bustos, O.~A. Zatarain~Durán, Adaptive {Control} for {Narrow}
  {Bandwidth} {Input} and {Output} {Disturbance} {Rejection} for a
  {Non}-{Isothermal} {CSTR} {System}, Mathematics 10~(18) (2022) 3224.
\newblock \href {https://doi.org/10.3390/math10183224}
  {\path{doi:10.3390/math10183224}}.

\bibitem{czyzniewski_observability_2022}
M.~Czyżniewski, R.~Łangowski, An observability and detectability analysis for
  non-linear uncertain {CSTR} model of biochemical processes, Scientific
  Reports 12~(1) (2022) 22327.
\newblock \href {https://doi.org/10.1038/s41598-022-26656-3}
  {\path{doi:10.1038/s41598-022-26656-3}}.

\bibitem{narendra_adaptive_2000}
K.~Narendra, O.~Driollet, Adaptive control using multiple models, switching,
  and tuning, in: Proceedings of the {IEEE} 2000 {Adaptive} {Systems} for
  {Signal} {Processing}, {Communications}, and {Control} {Symposium} ({Cat}.
  {No}.{00EX373}), IEEE, Lake Louise, Alta., Canada, 2000, pp. 159--164.
\newblock \href {https://doi.org/10.1109/ASSPCC.2000.882464}
  {\path{doi:10.1109/ASSPCC.2000.882464}}.

\end{thebibliography}

\appendix

\section{CTSR nonlinear functions}
\label{funciones}

The CTSR nonlinear functions are defined in equations \eqref{eqfnonlinear}-\eqref{eqhnonlinear}. The functions $f(\mathbf{x,u})$ are defined in the following way, according to the case study:

\subsection*{\textbf{Case 1: 1 reactor}} Let $f(\mathbf{x,u}) = [f_{1} \ f_{2} \ f_{3} \ f_{4} \ f_{5}]^{T} \in \mathbb{R}^{5\times 1}$ with
\begin{align}
f_{1} &= \frac{F_0 C_{A,0} - F_{i} C_{A,1} - k_i V C_{A,1} C_{B,1}}{V}, \\
f_{2} &= \frac{F_0 C_{B,0} - F_{i} C_{B,1} - k_i V C_{A,1} C_{B,1}}{V}, \\
f_{3} &= \frac{k_i V C_{A,1} C_{B,1} - F_{i} C_{C,1}}{V}, \\
f_{4} &= \frac{F_{i}}{V} (T_{\text{in},A} \!-\! T_1) - \frac{\Delta H}{\rho c_p} k_i C_{A,1} C_{B,1} + \frac{UA}{\rho c_p V} (T_{j,1} - T_1), \\
f_{5} &= \frac{F_j}{V_j} (T_{j,\text{in}} - T_{j,1}) - \frac{UA}{\rho_j c_{p_j} V_j} (T_{j,1} - T_1).
\end{align}

\subsection*{\textbf{Case 2: 2 reactors}} Let $f(\mathbf{x,u}) = [f_{1} \ f_{2} \ f_{3} \ \ldots \ f_{10}]^{T} \in \mathbb{R}^{10\times 1}$ with
\begin{align}
f_{6} &= \frac{F_{i} C_{A,1} - v C_{A,2} - k_i V C_{A,2} C_{B,2}}{V}, \\
f_{7} &= \frac{F_{i} C_{B,1} - v C_{B,2} - k_i V C_{A2} C_{B,2}}{V}, \\
f_{8} &= \frac{k_i V C_{A,2} C_{B,2} - F_{i} C_{C,2}}{V}, \\
f_{9} &= \frac{F_{i}}{V} (T_1 - T_2) - \frac{\Delta H}{\rho c_p} k_i C_{A,2} C_{B,2} + \frac{UA}{\rho c_p V} (T_{j,2} - T_2), \\
f_{10} &= \frac{F_j}{V_j} (T_{j,\text{in}} - T_{j,2}) - \frac{UA}{\rho_j c_{p_j} V_j} (T_{j,2} - T_2).
\end{align}

\subsection*{\textbf{Case 3: 3 reactors}} Let $f(\mathbf{x,u}) = [f_{1} \ f_{2} \ f_{3} \ \ldots \ f_{15}]^{T} \in \mathbb{R}^{15\times 1}$ with
\begin{align}
f_{11} &= \frac{F_{i} C_{A,2} - F_{i} C_{A,3} - k_i V C_{A,3} C_{B,3}}{V}, \\
f_{12} &= \frac{F_{i} C_{B,2} - F_{i} C_{B,3} - k_i V C_{A,3} C_{B,3}}{V}, \\
f_{13} &= \frac{k_i V C_{A,3} C_{B,3} - F_{i} C_{C,3}}{V}, \\
f_{14} &= \frac{F_{i}}{V} (T_2 - T_3) - \frac{\Delta H}{\rho c_p} k_i C_{A,3} C_{B,3} + \frac{UA}{\rho c_p V} (T_{j,3} - T_3), \\
f_{15} &= \frac{F_j}{V_j} (T_{j,\text{in}} - T_{j,3}) - \frac{UA}{\rho_j c_{p_j} V_j} (T_{j,3} - T_3).
\end{align}

\section{Extended Luenberger observer (ELO)}
\label{ELO}
The Luenberger observer estimates the state using the nonlinear dynamics of the system and a correction term based on the observation error:
\begin{equation}
\label{eq8}
\mathbf{\hat{\mathbf{x}}}_{k+1}=f(\mathbf{\hat{\mathbf{x}}}_{k},\mathbf{u}_{k}) + \mathbf{L} (\mathbf{y}_{k} - \hat{\mathbf{y}}_{k}),
\end{equation}
where $\mathbf{L}$ is the Luenberger observer gain, $\hat{\mathbf{y}}_{k} $ is the estimated output vector, and $\mathbf{y}_{k}$ is the measured output vector defined in Table \ref{tab_observability}. The Luenberger observer can be seen as a deterministic version of the Kalman filter when noise statistics are not considered.

\section{Extended Kalman Filter (EKF)}
\label{EKF}

The EKF operates recursively and constantly refines its estimate based on incoming measurements. It uses a local linearization of the nonlinear equations around the current operating point. It consists of two main stages: the prediction stage and the update stage.
\begin{enumerate}
\item Prediction Stage
\begin{align}\label{eq8b}
\hat{\mathbf{x}}_{k \mid k-1} &= f\left( \hat{\mathbf{x}}_{k-1 \mid k-1}, \mathbf{u}_{k-1} \right), \\
\hat{\mathbf{P}}_{k \mid k-1} &= \mathbf{F}_{k-1} \hat{\mathbf{P}}_{k-1 \mid k-1} \mathbf{F}_{k-1}^{T} + \mathbf{Q}.
\end{align}

\item Update Stage 
\begin{align}\label{eq8c}
\mathbf{K}_{k} &= \hat{\mathbf{P}}_{k \mid k-1} \mathbf{H}_{k}^{T} \left( \mathbf{H}_{k} \hat{\mathbf{P}}_{k \mid k-1} \mathbf{H}_{k}^{T} + \mathbf{R} \right)^{-1}, \\
\hat{\mathbf{x}}_{k \mid k} &= \hat{\mathbf{x}}_{k \mid k-1} + \mathbf{K}_{k} \left( \mathbf{y}_{k} - h(\hat{\mathbf{x}}_{k \mid k-1}) \right), \\
\hat{\mathbf{P}}_{k \mid k} &= \hat{\mathbf{P}}_{k \mid k-1} - \mathbf{K}_{k} \mathbf{H}_{k} \hat{\mathbf{P}}_{k \mid k-1},
\end{align}
\end{enumerate}
where
\begin{align}
\mathbf{F}_{k} = \left. \frac{\partial f(\mathbf{x}_k, \mathbf{u}_{k})}{\partial \mathbf{x}_k} \right|_{\mathbf{x}_k = \hat{\mathbf{x}}_{k\mid k}}, \quad
\mathbf{H}_{k} = \left. \frac{\partial h(\mathbf{x}_k)}{\partial \mathbf{x}_k} \right|_{\mathbf{x}_k = \hat{\mathbf{x}}_{k \mid k-1}},
\end{align}

\section{Unscented Kalman Filter (UKF)}
\label{UKF}

The UKF, for its part, uses the Unscented Transformation method for estimation. It generates a set of sigma points around the current estimate and propagates them through the nonlinear functions to more accurately capture system statistics. This provides a more accurate approximation for nonlinear systems without requiring derivatives. The general framework for UKF is based on the unscented transformation, in which a random variable $\mathbf{x} \sim \mathcal{N}(\mathbf{x}; \mathbf{m}, \bm{\Gamma})$, is transformed through a nonlinear function $\mathbf{y} = g(\mathbf{x})$. The basic idea is to choose a fixed number of sigma points that capture the mean and covariance of the original distribution of $\mathbf{x}$. The sigma points can be calculated using:
\begin{align}
\hat{\bm{\mathcal{X}}}_{i}&=\mathbf{m}, \ i=0 , \\  
\hat{\bm{\mathcal{X}}}_{i}&=\mathbf{m} \! + \! \sqrt{(n \! + \!\lambda)} \left [ \!\sqrt{\bm{\Gamma}} \right ]_{i}, \ i=1,\dots,L , \\ 
\hat{\bm{\mathcal{X}}}_{i}&=\mathbf{m} \! - \! \sqrt{(n \! + \!\lambda)} \left [ \!\sqrt{\bm{\Gamma}} \right ]_{i}, \ i=L+1,\dots,2L  ,
\end{align}
where $L$ is the length of $\mathbf{m}$. The scaling parameter $\lambda=\alpha ^{2}(L+\kappa)-L$ depends on the constant $\alpha$ and $\kappa$. $\alpha$ determines the spread of the sigma points around $\mathbf{m}$, and is usually set to $1\times 10^{-4}\leq \alpha \leq 1$. The constant $\kappa$ is usually set to 0. $ \left [ \sqrt{\bm{\Gamma}} \right ]_{i}$ is the $i_{th}$ column of the matrix $\sqrt{\bm{\Gamma}}$. For implementation, the parameters $\alpha$, $\beta$, and $\kappa$ are defined in Table \ref{tab_paramObserver}.

\begin{enumerate}
\item Prediction Stage: Form of the sigma points:
\begin{align}
\hat{\bm{\mathcal{X}}}_{i,k-1}&\!=\! \hat{\mathbf{x}}_{k-1}, \ i=0,  \\  
\hat{\bm{\mathcal{X}}}_{i,k-1}&\!=\! \hat{\mathbf{x}}_{k-1} \! + \! \sqrt{(n \! + \!\lambda)}  \left [ \!\!\sqrt{\mathbf{P}_{k-1}} \right ]_{i}, \ i \! = \! 1,\dots,L , \\ 
\hat{\bm{\mathcal{X}}}_{i,k-1}&\!=\! \hat{\mathbf{x}}_{k-1} \! - \! \sqrt{(n \! + \!\lambda)} \left [ \!\!\sqrt{\mathbf{P}_{k-1}} \right ]_{i}, \ i \!= \! L+1,\dots,2L  .
\end{align}

Propagate the sigma points through the dynamic model and $\hat{\mathbf{P}}_{k \mid k-1}$:
\begin{align}
\hat{\bm{\mathcal{X}}}_{i,k \mid k-1} & \! = \! f \left (\hat{\bm{\mathcal{X}}}_{i,k-1},\mathbf{u}_{k-1}  \right ), \ i=0,1,...,2L.
\end{align}

Compute the mean $\hat{\mathbf{x}}_{k \mid k-1}$ and the predicted covariance: 
\begin{align}
\hat{\mathbf{x}}_{k \mid k-1} & \! = \! \sum_{i=1}^{2L} W_{i}^{(m)}\hat{\bm{\mathcal{X}}}_{i,k \mid k-1}, \\
\hat{\mathbf{P}}_{k \mid k-1} & \! = \! \sum_{i=1}^{2L} W_{i}^{(c)}\left [ \hat{\bm{\mathcal{X}}}_{i,k \mid k-1} \!-\! \hat{\mathbf{x}}_{k \mid k-1} \right ]\left [ \hat{\bm{\mathcal{X}}}_{i,k \mid k-1} \!-\! \hat{\mathbf{x}}_{k \mid k-1} \right ]^{T} \!\!+\! \mathbf{Q} ,
\end{align}
where 
\begin{align}
W_{i}^{(c)} &= W_{i}^{(m)} = \frac{1}{\left [ 2(L+\lambda ) \right ]}, \ i=1,...2L,\\
W_{0}^{(c)} &=  \frac{\lambda}{(L+\lambda )}+(1-\alpha ^{2}+\beta ) , \ i=0, \\
W_{0}^{(m)} &=  \frac{\lambda}{(L+\lambda )} , \ i=0,
\end{align}

\item Update: Form of the sigma points:
\begin{align}
\hat{\bm{\mathcal{X}}}_{i,k }&=\hat{\mathbf{x}}_{k \mid k-1}, \ i=0,  \\  
\hat{\bm{\mathcal{X}}}_{i,k }&=\hat{\mathbf{x}}_{k \mid k-1} \! + \! \sqrt{(n \! + \!\lambda)}\left [ \!\!\sqrt{\mathbf{P}_{k \mid k-1}} \right ]_{i}, \ i=1,\dots,L  ,\\ 
\hat{\bm{\mathcal{X}}}_{i,k }&=\hat{\mathbf{x}}_{k \mid k-1} \! - \! \sqrt{(n \! + \!\lambda)}\left [ \!\!\sqrt{\mathbf{P}_{k \mid k-1}} \right ]_{i}, \ i=L+1,\dots,2L .
\end{align}

Propagate sigma points through the measurement model:
\begin{align}
\mathbf{z}_{i,k \mid k}&= h(\hat{\bm{\mathcal{X}}}_{i,k}), \  i=0,1,...,2L.
\end{align}

Compute the predicted mean $\hat{\mathbf{z}}_{k \mid k}$, the predicted covariance of the measurement $\mathbf{P}_{\mathbf{z}\mathbf{z}}$, and the cross-covariance of the state and the measurement $\mathbf{P}_{\mathbf{x}\mathbf{z}}$:
\begin{align}
\hat{\mathbf{z}}_{k \mid k}&= \sum_{i=0}^{2L}W_{i}^{(m)}\mathbf{z}_{i,k \mid k}, \\
\mathbf{P}_{\mathbf{z}\mathbf{z}} & \! = \! \sum_{i=0}^{2L}W_{i}^{(c)}\left [ \mathbf{z}_{i,k \mid k} \!-\! \hat{\mathbf{z}}_{k \mid k} \right ]\!\left [ \mathbf{z}_{i,k \mid k} \!-\! \hat{\mathbf{z}}_{k \mid k} \right ]^{T} \!\!+\! \mathbf{R}, \\ 
\mathbf{P}_{\mathbf{x}\mathbf{z}} & \! = \! \sum_{i=0}^{2L}W_{i}^{(c)} \!\left [ \hat{\bm{\mathcal{X}}}_{i,k \mid k} - \hat{\mathbf{x}}_{k \mid k} \right ] \left [ \mathbf{z}_{i,k \mid k} - \hat{\mathbf{z}}_{k \mid k} \right ]^{T}.
\end{align}

Compute the filter gain $\mathbf{K}_{k}$, the filtered state mean $\hat{\mathbf{x}}_{k \mid k}$ and the covariance $\hat{\mathbf{P}}_{k \mid k}$:
\begin{align}
\mathbf{K}_{k} & \! = \! \mathbf{P}_{\mathbf{x}\mathbf{z}}\mathbf{P}_{\mathbf{z}\mathbf{z}}^{-1},\\ 
\hat{\mathbf{x}}_{k \mid k}& \! = \! \hat{\mathbf{x}}_{k \mid k-1} +  \mathbf{K}_{k}\left ( \mathbf{y}_{k}  - \hat{\mathbf{z}}_{k \mid k}\right ), \\ 
\hat{\mathbf{P}}_{k \mid k}& \! = \! \hat{\mathbf{P}}_{k \mid k-1} -  \mathbf{K}_{k}\mathbf{P}_{\mathbf{z}\mathbf{z}}\mathbf{K}_{k}^{T}.
\end{align}
\end{enumerate}

\section{Quadrature Kalman Filter (QKF)}
\label{CKF}

The Quadrature Kalman Filter (QKF) uses the third-order quadrature principle to approximate integrals over Gaussian distributions (Gauss–Hermite numerical integration rule). This is achieved by evaluating the nonlinear functions at cubature points, specifically distributed, to capture the properties of the distributions. The QKF equations are divided into prediction and update stages, similar to other nonlinear Kalman filters such as the EKF and UKF. But unlike the UKF, where the sigma points are generated, the QKF presents an additional stage where the cubature points are generated and transformed. The number of dimensions $\mathcal{N}$, and the weights $\omega$ are defined in Table \ref{tab_paramObserver}.

\begin{enumerate} 
\item Prediction Stage

For a Gaussian distribution in $n$-dimensions $\mathcal{N}\left ( \hat{\mathbf{x}}_{k-1},\hat{\mathbf{P}}_{k-1} \right )$ with the associated weights $\omega _{l=1}^{n}$ the cubature points are:
\begin{equation}  
\mathcal{\bm{\mathcal{X}}}_{i,k-1}=\hat{\mathbf{x}}_{k-1}+\sqrt{n\hat{\mathbf{P}}_{k-1}}\xi_{i}, \ i =1,...,2n,
\end{equation}
where $\sqrt{n\hat{\mathbf{P}}_{k-1}}$ is the weighted square root of the covariance matrix scaled by $n$, and $\xi_{i}$ are Gauss–Hermite quadrature points with weights $\omega _{l}$.

Propagate the cubature points through the dynamic model:
\begin{equation}
\bm{\mathcal{X}}_{i,k\mid k-1}^{*}=f(\bm{\mathcal{X}}_{i,k-1},\mathbf{u}_{k-1}) \ i =1,...,n,
\end{equation}
where $f(\mathcal{\bm{\mathcal{X}}}_{i,k-1},\mathbf{u}_{k-1})$ is the nonlinear function that describes the dynamics of the system.

Compute the mean and the predicted covariance:
\begin{align}
\hat{\mathbf{x}}_{k \mid k-1} & \! = \! \dfrac{1}{2n} \sum_{i=1}^{2n} \bm{\mathcal{X}}_{i,k \mid k-1}^{*}. \\ 
\hat{\mathbf{P}}_{k \mid k-1} & \! = \! \dfrac{1}{2n}\sum_{i=1}^{2n}\left [ \bm{\mathcal{X}}_{i,k \mid k-1}^{*} \!-\! \hat{\mathbf{x}}_{k \mid k-1} \right ]\left [ \bm{\mathcal{X}}_{i,k \mid k-1 }^{*} \!-\! \hat{\mathbf{x}}_{k \mid k-1} \right ]^{T} \!+\! \mathbf{Q}.
\end{align}
\item Update

Form the cubature points:
\begin{align}
\bm{\mathcal{X}}_{i,k}^{*} & = \hat{\mathbf{x}}_{k \mid k-1} + \sqrt{n\hat{\mathbf{P}}_{k \mid k-1}}\xi_{i}, \ i =1,...,2n.
\end{align}

Propagate sigma points through the measurement model:
\begin{align}
\mathbf{z}_{i,k \mid k} & = h(\bm{\mathcal{X}}_{i,k}^{*}) \ i=1,...,2n.
\end{align}

Compute the predicted mean, the predicted covariance of the
measurement, and the cross-covariance of the state and the measurement
\begin{align}
\hat{\mathbf{z}}_{k \mid k} &= \frac{1}{2n}\sum_{i=1}^{2n}\mathbf{z}_{i,k \mid k}, \\ 
\hat{\mathbf{P}}_{\mathbf{z}\mathbf{z}} &= \frac{1}{2n}\sum_{i=1}^{2n}\left [ \mathbf{z}_{i,k \mid k} - \hat{\mathbf{z}}_{k \mid k} \right ]\left [ \mathbf{z}_{i,k \mid k} - \hat{\mathbf{z}}_{k \mid k} \right ]^{T} + \mathbf{R},\\ 
\hat{\mathbf{P}}_{\mathbf{x}\mathbf{z}} &= \frac{1}{2n}\sum_{i=1}^{2n}\left [ \bm{\mathcal{X}}_{i,k \mid k}^{*} -\hat{\mathbf{x}}_{k\mid k} \right ]\left [ \mathbf{z}_{i,k \mid k} - \hat{\mathbf{z}}_{k \mid k} \right ]^{T}.
\end{align}

Compute the filter gain $\mathbf{K}_{k}$, the filtered state mean $\hat{\mathbf{x}}_{k \mid k}$ and the covariance $\hat{\mathbf{P}}_{k \mid k}$:
\begin{align}
\mathbf{K}_{k} &= \mathbf{P}_{\mathbf{x}\mathbf{z}}P_{\mathbf{z}\mathbf{z}}^{-1}, \\
\hat{\mathbf{x}}_{k\mid k}&= \hat{\mathbf{x}}_{k \mid k-1} +  \mathbf{K}_{k}\left ( \mathbf{y}_{k}  - \hat{\mathbf{z}}_{k \mid k}\right ), \\
\hat{\mathbf{P}}_{k\mid k}&= \hat{\mathbf{P}}_{k \mid k-1} -  \mathbf{K}_{k}\mathbf{P}_{\mathbf{z}\mathbf{z}}\mathbf{K}_{k}^{T}.
\end{align}

\end{enumerate}

\section{Particle Filter (PF)}
\label{PF}

A particle filter is a simulation-based estimation method that uses a set of samples or "particles" to approximate the probability distribution of a nonlinear and/or non-Gaussian system. The Particle Filter (PF) shares several similarities with the EKF, UKF, and QKF. Its approach also has key differences, requiring no linearization or sigma points; instead, it uses particle-based simulations. The number of particles $N$ is defined in Table \ref{tab_paramObserver}.

The main steps that describe its operation are presented below:
\begin{enumerate}
\item Particle initialization Stage: Draw $N$ particles $\mathbf{x}_{i,0} \sim p(\hat{\mathbf{x}}_{1})$ and set $w_{i,0}=1/N$ for $i=1,2,…,N$.
\item Draw samples from the importance distributions
\begin{align}
    \mathbf{x}_{i,k} & \sim p(\mathbf{x}_{k}\mid \mathbf{x}_{i,k-1}),  \qquad i=1,…,N .
\end{align}
\item Calculate new weights according to:
\begin{align}
    w_{i,k} \propto p(\mathbf{y}_{k}\mid \mathbf{x}_{i,k}),  \qquad i=1,…,N ,
\end{align}
and normalize them to sum to unity. The state estimation $\hat{\mathbf{x}}{k\mid k}$ and its corresponding estimation error covariance $\mathbf{P}{k\mid k}$ are performed as the weighted average of the particles:
\begin{align}
    \hat{\mathbf{x}}_{k\mid k} &=  \sum_{i=1}^{N} w_{i,k}\mathbf{x}_{i,k},\\
    \mathbf{P}_{k\mid k} &=  \sum_{i=1}^{N} w_{i,k}(\mathbf{x}_{i,k}-\hat{\mathbf{x}}_{k\mid k})(\mathbf{x}_{i,k}-\hat{\mathbf{x}}_{k\mid k})^{T}.
\end{align}

\item Do resampling.
 
\end{enumerate}
\begin{figure}[H]
	\centering
    {\includegraphics[width=\columnwidth]{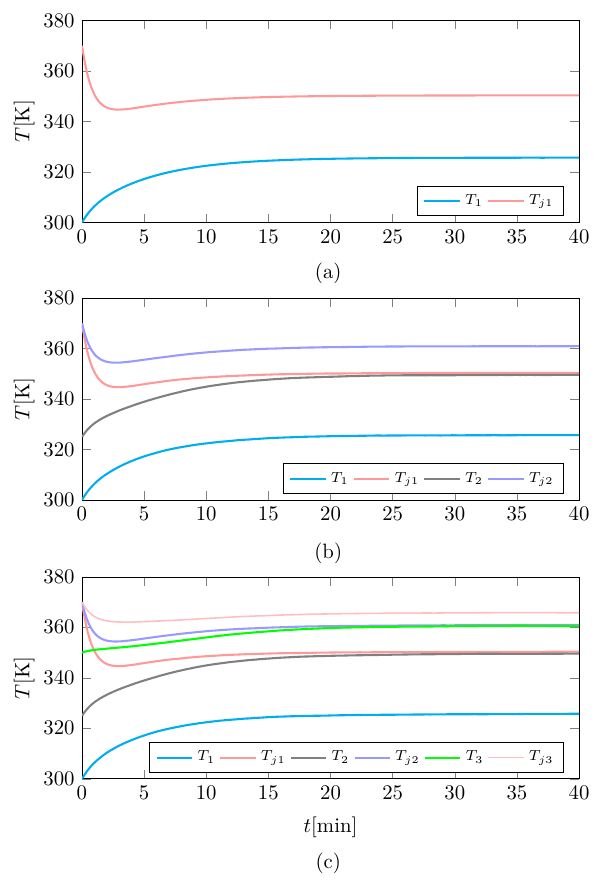}}
	\centering
    \caption{Temperature measurements available for state estimation in each case: (a) Case 1: internal ($T_1$, cyan) and jacket ($T_{j1}$, light red) temperatures of Reactor 1; (b) Case 2: internal ($T_1$, cyan; $T_2$, gray) and jacket ($T_{j1}$, light red; $T_{j2}$, light purple) temperatures of Reactors 1 and 2; (c) Case 3: internal ($T_1$, cyan; $T_2$, gray; $T_3$, green) and jacket ($T_{j1}$, light red; $T_{j2}$, light purple; $T_{j3}$, pink) temperatures of Reactors 1, 2, and 3.}
\label{fig:mediciones_Temp}
\end{figure}
\section{Temperature}
\label{App_T}

The time evolution of the measurable temperatures obtained by simulating the CSTR system with the nonlinear model is shown in Figure \ref{fig:mediciones_Temp}. These correspond to the $\mathbf{y}$ vectors (Table \ref{tab_observability}) used for state estimation across all designed observers.

\end{document}